\documentclass[journal]{IEEEtran}
\usepackage{graphicx}
\usepackage{epstopdf}
\usepackage{subfigure}
\usepackage{color}

\ifCLASSINFOpdf
\else
\fi
\begin{document}
%
\title{A Self-Learning Information Diffusion Model for Smart Social Networks}
%
%
%

\author{Qi~Xuan,~\IEEEmembership{Member,~IEEE},
        Xincheng~Shu,
        Zhongyuan~Ruan,
        Jinbao~Wang,
        Chenbo~Fu,
        and~Guanrong~Chen,~\IEEEmembership{Fellow,~IEEE}
\thanks{Q. Xuan, X. Shu, J. Wang, and C. Fu are with the College of Information Engineering, Zhejiang University of Technology, Hangzhou 310023, China (e-mail:xuanqi@zjut.edu.cn; sxc.shuxincheng@foxmail.com; Jinbaowang\_zjut@yeah.net; cbfu@zjut.edu.cn).}
\thanks{Z. Ruan is with the College of Computer Science and Technology, Zhejiang University of Technology, Hangzhou 310023, China (e-mail:
zyruan@zjut.edu.cn).}
\thanks{G. Chen is with the Department of Electronic Engineering, City University of Hong Kong, Hong Kong, China (e-mail: eegchen@cityu.edu.hk).}
\thanks{This article has been submitted on Nov 4th, 2018.}
\thanks{Corresponding author: Zhongyuan Ruan.}}
%
%

\markboth{}%
{Shell \MakeLowercase{\textit{et al.}}: Bare Demo of IEEEtran.cls for IEEE Journals}
%



\maketitle

\begin{abstract}
In this big data era, more and more social activities are digitized thereby becoming traceable, and thus the studies of social networks attract increasing attention from academia. It is widely believed that social networks play important role in the process of information diffusion. However, the opposite question, i.e., how does information diffusion process rebuild social networks, has been largely ignored. In this paper, we propose a new framework for understanding this reversing effect. Specifically, we first introduce a novel information diffusion model on social networks, by considering two types of individuals, i.e., \emph{smart} and \emph{normal} individuals, and two kinds of messages, \emph{true} and \emph{false} messages. Since social networks consist of human individuals, who have \emph{self-learning} ability, in such a way that the trust of an individual to one of its neighbors increases (or decreases) if this individual received a true (or false) message from that neighbor. Based on such a simple self-learning mechanism, we prove that a social network can indeed become smarter, in terms of better distinguishing the true message from the false one. Moreover, we observe the emergence of \emph{social stratification} based on the new model, i.e., the true messages initially posted by an individual closer to the smart one can be forwarded by more others, which is enhanced by the self-learning mechanism. We also find the \emph{crossover advantage}, i.e., interconnection between two chain networks can make the related individuals possessing higher social influences, i.e., their messages can be forwarded by relatively more others. We obtained these results theoretically and validated them by simulations, which help better understand the reciprocity between social networks and information diffusion.
\end{abstract}

%
\IEEEpeerreviewmaketitle

\section{Introduction}
\IEEEPARstart{S}{ocial} networks~\cite{kossinets2006empirical,borgatti2009network} have been extensively studied in recent years, partly due to the availability of big electronic communication data from multi-media such as phone calls~\cite{xuan2009empirical}, emails~\cite{newman2002email}, tweets~\cite{vespignani2012modelling}, etc. Many studies focused on analyzing the structures of social networks. Barab\'{a}si et al.~\cite{barabasi1999emergence} established a movie actor collaboration network which has a power-law degree distribution, referred to as a scale-free network. Xuan et al.~\cite{xuan2009empirical} performed an empirical analysis on the Internet telephone network and established an ID-to-phone bipartite communication network. They found that the network has a hierarchical and modular structure, and most of the weak links connect to the ID nodes of large degrees in the giant component, indicating the important roles of weak links in keeping the structure of the network. Myers et al.~\cite{myers2014information} differentiate  social networks from information networks, defining a social network by high degree assortativity, small shortest path length, large connected component, high clustering coefficient, and high degree of reciprocity, while defining an information network by large node degrees, lack of reciprocity, and large two-hop neighborhoods. Based on these definitions, they found that, from an individual user's perspective, Twitter starts more like an information network, but evolves to behave more like a social network.

Besides network structures, it is also widely recognized that social networks play significant roles in many social processes~\cite{kleinberg2000navigation,dodds2003information,xuan2013synchrony,centola2010spread,iniguez2017service}, especially information diffusion. Recently, Cha et al.~\cite{cha2009measurement} collected and analyzed large-scale traces of information dissemination in the Flickr social network. They found that even popular photos spread slowly and narrowly throughout the network, but the information exchanged between friends seems to account for over 50\% of all favorite markings. Yang et al.~\cite{yang2010understanding} studied the retweeting behaviors and found that almost 25.5\% of the tweets posted by users are actually retweeted from friends' blog spaces. They further proposed a factor graph model to predict users' retweeting behaviors, achieving a precision of 28.81\% and a recall of 37.33\%. Furthermore, Myers et al.~\cite{myers2012information} presented a model in which information can reach a node via the links on a social network or through the influence of external sources. They found that about 71\% of the information volume in Twitter are attributed to network diffusion, while only 29\% is due to external events and factors outside the network. Liu et al.~\cite{liu2015events} analyzed the diffusion of eight typical events  on Sina Weibo,  and found that external influence indeed has significant impact on information spreading, confirming the out-of-social-network influence.

Different nodes and links in a network may play quite different roles in information diffusion. Kitsak et al.~\cite{kitsak2010identification} applied the susceptible-infectious-recovered (SIR) and susceptible-infectious-susceptible (SIS) models~\cite{pastor2015epidemic} on several real-world complex networks, and found that the most efficient spreaders are those located within the core of the network as identified by the k-shell decomposition analysis~\cite{carmi2007model}. They also found that, when multiple spreaders are involved simultaneously, the average distance among them becomes the crucial parameter that determines the extent of the spreading.  L{\"u} et al.~\cite{lu2016h} constructed an operator which strings together some widely used metrics for identifying influential nodes, including degree, H-index and coreness. Their analyses showed that the H-index in many cases can better quantify node influence than degree and coreness. Bakshy et al.~\cite{bakshy2012role} investigated the roles of strong and weak links in information propagation by designing experiments on Facebook, showing that stronger links are more influential individually, while the larger number of weak links are responsible for the propagation of novel information. Most of these works focus on the structural differences among nodes or links and then study their effects on information diffusion, but largely ignore some essential differences among social individuals, e.g., some individuals might be smarter than the others in distinguishing rumors.

It has been noticed that information of different topics might have different diffusion patterns. Romero et al.~\cite{romero2011differences} studied the differences of information diffusion mechanisms across many topics on Twitters. They found that the hashtags on politically controversial topics are relatively persistent, i.e., repeated exposures continue to have unusually large marginal effects on adoption; while the hashtags representing the natural analogues of Twitter idioms and neologisms are relatively non-persistent, i.e., the effect of multiple exposures decays rapidly comparing with the first exposure. Xuan et al.~\cite{xuan2013reaction} proposed an SIS model on duplex networks, considering that different layered networks transfer different kinds of data. Based on that model, they found that spreading may be promoted or suppressed if the degree sequences of a two layered network are positively or negatively correlated, respectively.

There are also many works about rumor spreading on social networks. Zhou et al.~\cite{zhou2007influence} considered the influence of network topological structure and the unequal footings of neighbors of an infected node on propagating rumors, and found that the final infected density decreases as the structure changes from random to scale-free network. Fountoulakis et al.~\cite{fountoulakis2012ultra} adopted a push-pull protocol to study rumor spreading, and theoretically proved that a rumor spreads very fast from one node to all others, i.e., within $O(\log \log n)$ rounds, for a random network that has a power-law degree distribution with an exponent between 2 and 3. Wang et al.~\cite{ya2013rumor} introduced a trust mechanism into the SIR model, and found that the trust mechanism greatly reduces the maximum rumor influence, and meanwhile it postpones the rumor terminal time, providing a better opportunity to control the rumor spreading. Trpevski et al.~\cite{trpevski2010model} generalized the SIS model by considering two rumors with one prior to the other. They found that the preferred rumor is dominant in the network when the degrees of nodes are high enough and/or when the network contains large clustered groups of nodes, but it seems to be also possible for the other rumor to occupy some fraction of the nodes as well. Huang and Jin~\cite{huang2011preventing} further applied the random and targeted immunization strategies to the rumor model on a small-world network, and found that both strategies are effective in spreading rumors when the average degree of the network is small, but they will lose efficiency when the average degree is large. In the area of online rating systems, there are a lot of fake reviews, e.g., roughly 16\% of restaurant reviews on Yelp are filtered~\cite{luca2016fake}, which tend to be more extreme than the others. Luca et al.~\cite{luca2016fake} then revealed the economic incentives behind a business decision to leave fake reviews: independent restaurants are more likely to leave positive fake reviews for themselves, but negative fake reviews are more likely to occur when a business has an independent competitor. These empirical results on rumors and fake reviews suggest considering both true and false messages in information diffusion models.

On the other hand, there are several studies that focus on understanding the reverse effects of information diffusion on network evolution. Weng et al.~\cite{weng2013role} studied the evolution of a social network on Yahoo! Meme. They found that, while triadic closure is the dominant mechanism for social network evolution in the early stages of a user's lifetime, the traffic generated by the dynamics of information flow on the network becomes an indispensable component for user linking behavior as time progresses. Those users who are popular, active, and influential tend to create traffic-based shortcuts, making the information diffusion process more efficient over the network. Xuan et al.~\cite{xuan2010emergence} suggested that reaction-diffusion (RD) processes, rather than pure topological rules, might be responsible for the emergence of heterogeneous structures of complex networks. They further proposed a framework for controlling the RD process by adjusting the structure of the underlying diffusion network~\cite{xuan2011structural}. These results suggest integrating a learning mechanism into information diffusion models, which can make the social networks smarter as time evolves, i.e., tending to amplify true messages or diminish false messages.

In this paper, we aim to establish a theoretical self-learning model to study the information diffusion on a network with different types of nodes and different kinds of messages. This model and other main contributions of this paper are summarized as follows:
\begin{itemize}
  \item First, we introduce a new information diffusion model by considering two types of nodes, i.e., \emph{smart} nodes and \emph{normal} nodes, and two different kinds of messages, \emph{true} messages and \emph{false} messages.
  \item Second, we propose a metric to measure the \emph{information filtering ability} (IFA) of a social network, which is defined as the relative difference between the spreading ranges of true message and false message.
  \item Third, we integrate a \emph{self-learning mechanism} into the model, with the trust among social individuals evolves with time, i.e., the trust of an individual to one of its neighbors increases/decreases if this individual received a true/false message from that neighbor. This can make the social network as a whole gradually becoming smarter.
      \item Finally, we theoretically and numerically analyze the self-learning model, and investigate the emergence of two basic social concepts, i.e., \emph{social stratification}~\cite{saunders2006social} and \emph{crossover advantage}~\cite{henni2015crossover}, within the new framework.
\end{itemize}

The rest of the paper is organized as follows. In Section~\ref{model}, we propose a self-learning information diffusion model by considering two types of nodes and two kinds of messages. Meanwhile, we define the information filtering ability of a social network. In Section~\ref{TA}, we perform the theoretical analysis of the model on chain and star networks, and study the emergence of social stratification and crossover advantage. In Section~\ref{NR}, we validate the theoretical results by simulations. We conclude the investigation by Section~\ref{Conclusion}.

\section{Self-Learning Information Diffusion Model\label{model}}
Differing from the traditional information diffusion model on networks, here we suppose that there are two different kinds of messages on a network, i.e., true and false messages, represented by 1 and 0 respectively. And we also suppose that there are two different types of nodes, i.e., smart nodes that can precisely distinguish whether a message is true or false, and normal nodes that cannot do so.

\subsection{Basic Cascading Model\label{BCM}}
Assume that diffusion occurs on a weighted directed network represented by a graph $G=(V,E,W)$ with nodes $V=\{v_1,v_2,\ldots,v_N\}$ and links $E\subset{V\times{V}}$. Each directed link $e_{ij}\in{E}$ has a weight $w_{ij}\in{W}$, which is a real number and satisfies $0\leq{w_{ij}}\leq{1}$, and is used to measure the trust of $v_j$ to $v_i$.  Denote by $S$ and $O$ the sets of smart nodes and normal nodes, respectively, satisfying $S\cup{O}=V$ and $S\cap{O}=\emptyset$. Based on these definitions, we propose a new cascading model as follows.
\begin{enumerate}
  \item \textbf{Assignment}. A subset $S$ of nodes are selected as smart nodes, while the rest are normal nodes in $O$.
  \item \textbf{Triggering}. A node $v_j$ is randomly chosen as the source node, which sends out a true message if it is a smart node, i.e., $v_j\in{S}$, and sends out a true or false message with an equal probability if it is a normal node, i.e., $v_j\in{O}$.
  \item \textbf{Cascading}. When a node $v_k$ received a message from its incoming neighbors, it will first randomly pick one of these incoming neighbors, denoted by $j$, to follow. Then, if $v_k$ is a smart node, i.e., $v_k\in{S}$, it will forward the message with probability $p=\eta$ if it is true and decline to transmit it otherwise. If $v_k$ is a normal node, i.e., $v_k\in{O}$, it will forward the message with probability $p=\eta{w_{jk}}$ no matter whether it is true or false. Here, $0\leq\eta\leq{1}$ is the \emph{natural forwarding rate} (NFR). This is because, as defined, a normal node cannot distinguish the true from the false, and thus it will copy the behavior of a close incoming neighbor to follow, with the probability proportional to the trust defined by the directed weight from $v_j$ to $v_k$. Assume that if a node received but denied to transmit the message, it will never forward it in the future.
\end{enumerate}

Based on this model, after cascading, we can count the numbers of nodes that deliver true messages and false messages, denoted by $N_T$ and $N_F$, respectively.  Then, we define the \emph{true message transmission ability} (TTA) and \emph{false message transmission ability} (FTA) as
\begin{eqnarray}
     F_T&=&\frac{N_T}{N},\\ \label{Eq:TTA}
     F_F&=&\frac{N_F}{N},\label{Eq:FIA}
\end{eqnarray}
respectively, based on which we can further define the \emph{information filtering ability} (IFA) of a social network, as
\begin{equation}
     F=\frac{F_T-F_F}{F_F}.\label{Eq:IFA}
\end{equation}
Generally, a larger value of $F$ indicates a relatively stronger IFA of a network, i.e., it can promote the true message or downgrade the false message; in other words, it can deliver true (or false) messages to more (or fewer) nodes.

\subsection{Self-Learning Mechanism\label{SLM}}
In a real social network, the trust among people may evolve with time. Imaging the following scene: an individual received a message from a friend. Suppose that this individual can finally be notified whether this message is true or false. Then, intuitively this person will trust the friend more if the message is true but less if the message is false. In this study, we aim to simulate such reward and punishment mechanism through a re-weighting process in a social network.

The self-learning mechanism thus consists of the following three steps.
\begin{enumerate}
  \item \textbf{Initialization}. Each directed link from $v_j$ to $v_k$ in the network is assigned with a constant value as its weight, i.e., $w_{jk}=0.5$, to represent the trust of $v_k$ to $v_j$.
  \item \textbf{Re-weighting process}. For each iteration $t$, the \emph{triggering} and \emph{cascading} steps in the basic cascading model are performed. Suppose there is a directed link from $v_j$ to $v_k$. If the message was delivered from $v_j$ to $v_k$, then update the link weight by
\begin{equation}
w_{jk}(t+1)=
\left\{
\begin{array}{ll}
w_{jk}(t)+\Delta & w_{jk}(t)\leq{1-\Delta}\\
1 & w_{jk}(t)>{1-\Delta}
\end{array}\right.
\label{Eq:WTrue}
\end{equation}
if it is true; and by
\begin{equation}
w_{jk}(t+1)=
\left\{
\begin{array}{ll}
w_{jk}(t)-\Delta & w_{jk}(t)\geq{\Delta+\delta}\\
\delta & w_{jk}(t)<\Delta+\delta
\end{array}\right.
\label{Eq:WFalse}
\end{equation}
if it is false. Here, $\Delta$ is a constant of relatively small positive value, representing the reward or punishment in each round; and $\delta$ is also a constant of small positive value, which is used to avoid zero transition probability between each pair of linked nodes. In this study, both of them are set to 0.01. Note that if $v_k$ received the message from $v_j$ but denied to deliver it, the link weight will also be updated by Eq.~(\ref{Eq:WTrue}) if the message is true and by Eq.~(\ref{Eq:WFalse}) if it is false.
\item \textbf{Termination}. When the link weights in the network are relatively stable or the number of iterations reaches $M$, the whole re-weighting process is terminated.
\end{enumerate}

\section{Theoretical Analysis\label{TA}}
Now, the above model is analyzed theoretically, and the link weights in the network are estimated, and further the three abilities, i.e., TTA, FTA, and IFA, are calculated. In particular, chain and star networks are discussed for simplicity, which are two of the most basic building blocks or motifs of many real-world social networks.

\subsection{Chain Network\label{chain}}
\begin{figure}[!h]
\centering
\includegraphics[width=\linewidth]{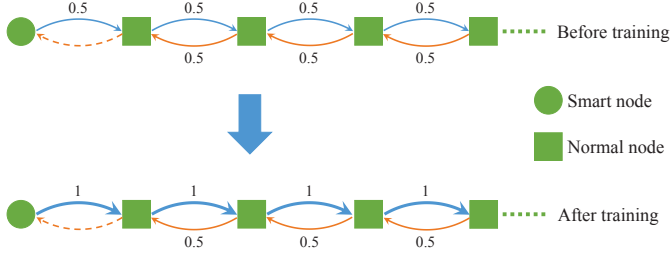}\caption{A chain network with one terminal being the only smart node, before and after training. Note that the direcred link from a normal node to a smart node will not influence the decision of the smart node to forward a message or not. Thus, a dashed directed line is used to mean that the smart node can find messages from the normal nodes.}
\label{ChainNetwork}
\end{figure}

Given a chain network, with one terminal being a smart node, the rest being normal nodes, and  the directed weights between pairwise-connected nodes set to 0.5 initially, as shown in Fig.~\ref{ChainNetwork}. To calculate the three abilities, i.e., TTA, FTA, and IFA, of this chain network without self-learning, assume that the smart node can endow the network with a higher ability to distinguish the true messages from the false, but such ability might diminish quickly as the length of the chain increases.

\emph{\textbf{Theorem 1:}} The IFA of the chain network with one terminal being a smart node is always positive, indicating that the smart node can enable the network to distinguish the true messages from the false. And such an ability is enhanced by increasing the NFR $\eta$ or decreasing the network size $N$.

\emph{\textbf{Proof:}} Without loss of generality, denote the smart node as $v_1$, and the other nodes with increasing indexes from the smart node, i.e., $v_{i}$ is connected to $v_{i+1}$. Suppose $v_i$ is the source node which posts a true message. The number of nodes in the network that consequently post the message can be estimated by
\begin{equation}
n_T(i)=\sum_{k=0}^{N-i}\left(\frac{\eta}{2}\right)^k+\sum_{k=1}^{i-2}\left(\frac{\eta}{2}\right)^k+\eta\left(\frac{\eta}{2}\right)^{i-2},
\label{Eq:CNT1}
\end{equation}
when $i\geq{2}$, and
\begin{equation}
n_T(i)=\sum_{k=0}^{N-1}\left(\frac{\eta}{2}\right)^k,
\label{Eq:CNT2}
\end{equation}
when $i=1$. For Eq.~(\ref{Eq:CNT1}),  the first term estimates the number of nodes that consequently post the message in the subnetwork from $v_i$ to $v_N$; the second term estimates that in the subnetwork from $v_2$ to $v_{i-1}$; and the third term is the probability that the smart node $v_1$ posts the message. Suppose each node has an equal probability to be selected as the source node. Then, the average TTA of the network can be calculated by
\begin{eqnarray}
F_T&=&\frac{1}{N}\left[\frac{n_T(1)}{N}+\frac{\sum_{i=2}^N{n_T(i)}}{N}\right]\nonumber\\
&=&\frac{1}{N^2}\frac{1-(\eta/2)^N}{1-\eta/2}\nonumber\\
&+&\frac{1}{N^2}\left[\frac{N-1}{1-\eta/2}-\frac{\eta}{2}\frac{1-(\eta/2)^{N-1}}{(1-\eta/2)^2}\right]\nonumber\\
&+&\frac{1}{N^2}\left[\frac{\eta}{2}\frac{N-1}{1-\eta/2}-\frac{\eta}{2}\frac{1-(\eta/2)^{N-1}}{(1-\eta/2)^2}\right]\nonumber\\
&+&\frac{\eta}{N^2}\frac{1-(\eta/2)^{N-1}}{1-\eta/2}.
\label{Eq:FTChain}
\end{eqnarray}
As $N\to\infty$, one has $(\eta/2)^{N-1}\to{0}$, since $\eta/2$ must be smaller than 1. In this case, Eq.~(\ref{Eq:FTChain}) can be simplified to
\begin{equation}
F_T=\frac{1}{N^2}\left[\frac{1+\eta}{1-\eta/2}N-\frac{\eta}{(1-\eta/2)^2}\right].
\label{Eq:FTChainSimple}
\end{equation}

Now, suppose $v_i$ as the source node posts a false message. The number of nodes in the network that consequently post this false message can be estimated by
\begin{equation}
n_F(i)=\sum_{k=1}^{i-2}\left(\frac{\eta}{2}\right)^k+\sum_{k=0}^{N-i}\left(\frac{\eta}{2}\right)^k,
\label{Eq:CNF}
\end{equation}
when $i\geq{2}$. It should be noted that $v_1$ as the smart node will never post any false message. Similarly, the average FTA of the network can be calculated by
\begin{eqnarray}
F_F&=&\frac{\sum_{i=2}^N{n_F(i)}}{N(N-1)}\nonumber\\
&=&\frac{1}{N(N-1)}\left[\frac{\eta}{2}\frac{N-1}{1-\eta/2}-\frac{\eta}{2}\frac{1-(\eta/2)^{N-1}}{(1-\eta/2)^2}\right]\nonumber\\
&+&\frac{1}{N(N-1)}\left[\frac{N-1}{1-\eta/2}-\frac{\eta}{2}\frac{1-(\eta/2)^{N-1}}{(1-\eta/2)^2}\right],
\label{Eq:FFChain}
\end{eqnarray}
which can be simplified to
\begin{equation}
F_F=\frac{1}{N(N-1)}\left[\frac{1+\eta}{1-\eta/2}(N-1)-\frac{\eta}{(1-\eta/2)^2}\right],
\label{Eq:FFChainSimple}
\end{equation}
when $N$ is large enough.

Based on Eqs.~(\ref{Eq:IFA}), (\ref{Eq:FTChainSimple}) and (\ref{Eq:FFChainSimple}), one can estimate the IFA as follows:
\begin{eqnarray}
F&=&\frac{F_T-F_F}{F_F}\nonumber\\
&=&\frac{\eta}{N\left[(1+\eta)(1-\eta/2)(N-1)-\eta\right]}\nonumber\\
&\sim&\frac{\eta}{N^2(1+\eta)(1-\eta/2)}.
\label{Eq:IFAChain}
\end{eqnarray}
Eq.~(\ref{Eq:IFAChain}) indicates that the IFA of the chain network with one terminal being a smart node is always positive, and it is an increasing function of the NFR $\eta$ but a decreasing function of the network size $N$. This completes the proof.

\emph{\textbf{Remark 1:}} Now, the network will be trained based on the self-learning mechanism introduced in Sec.~\ref{SLM}. Note that, if a normal node is selected as the source node, it will post a true or false message with an equal probability, while if the smart node is selected as the source node, it will post a true message with probability $\eta$ but never post a false message. When the smart node is not selected as the source node, node $v_i$ will always observe the true or false message with an equal probability from any of its neighbors, i.e., $v_{i-1}$ and $v_{i+1}$. In this case,  based on the re-weighting process, the weight of the directed link from $v_{i-1}$ (or $v_{i+1}$) to $v_i$ will be close to its original weight 0.5, on the average. However, in the present model, the smart node $v_1$ can be selected as the source node. In this case, the true message it posts could be delivered to $v_{i-1}$, which could be further observed by $v_i$, and thus the directed weight from $v_{i-1}$ to $v_i$ may increase, i.e., the weight can be considered as an increasing function of the number of iterations statistically. Since the probability that $v_{i-1}$ forwards the message initially posted by the smart node is determined by the shortest directed path length from $v_1$ to $v_{i-1}$ and the weights on the associated directed links, and the increment $\Delta$ is a constant, all the weights of the directed links from $v_{i-1}$ to $v_i$, for $i=2,3,\ldots,N$, should tend to be 1 when the number of iterations is large enough. Note that, based on the self-learning mechanism, the extra true messages posted by the smart node will not influence the weights of the directed links from $v_{i}$ to $v_{i-1}$. Therefore, after sufficiently many iterations, one can get a network with all the weights of the directed links from $v_{i-1}$ to $v_i$ equal to 1, while most of the weights of the directed links from $v_{i}$ to $v_{i-1}$ close to 0.5, for $i=2,3,\ldots,N$, as shown in Fig.~\ref{ChainNetwork}.

\emph{\textbf{Theorem 2:}} After training, the IFA of the chain network with one terminal being a smart node is still positive, and is even larger than that before training, indicating that the self-learning mechanism can enable the chain network to become smarter, in the sense of better distinguishing a true message from the false.

\emph{\textbf{Proof:}} For the re-weighted chain network, suppose $v_i$ is the source node which posts a true message. The number of nodes in the network that consequently post this true message can be estimated by
\begin{equation}
n_T(i)=\sum_{k=0}^{N-i}\eta^k+\sum_{k=1}^{i-2}\left(\frac{\eta}{2}\right)^k+\eta\left(\frac{\eta}{2}\right)^{i-2},
\label{Eq:CNTTrain1}
\end{equation}
when $i\geq{2}$, and
\begin{equation}
n_T(i)=\sum_{k=0}^{N-1}\eta^k,
\label{Eq:CNTTrain2}
\end{equation}
when $i=1$. In this case, the average TTA of the network can be calculated by
\begin{eqnarray}
F_T&=&\frac{1}{N}\left[\frac{n_T(1)}{N}+\frac{\sum_{i=2}^N{n_T(i)}}{N}\right]\nonumber\\
&=&\frac{1}{N^2}\frac{1-\eta^N}{1-\eta}+\frac{1}{N^2}\left[\frac{N-1}{1-\eta}-\eta\frac{1-\eta^{N-1}}{(1-\eta)^2}\right]\nonumber\\
&+&\frac{1}{N^2}\left[\frac{\eta}{2}\frac{N-1}{1-\eta/2}-\frac{\eta}{2}\frac{1-(\eta/2)^{N-1}}{(1-\eta/2)^2}\right]\nonumber\\
&+&\frac{\eta}{N^2}\frac{1-(\eta/2)^{N-1}}{1-\eta/2}.
\label{Eq:FTChainTrain}
\end{eqnarray}
As $N\to\infty$, Eq.~(\ref{Eq:FTChainTrain}) can be simplified to
\begin{eqnarray}
F_T&=&\frac{1}{N}\frac{2+\eta-2\eta^2}{(2-\eta)(1-\eta)}\nonumber\\
&-&\frac{1}{N^2}\left[\frac{2\eta}{(2-\eta)^2}+\frac{\eta}{(1-\eta)^2}\right].
\label{Eq:FTChainSTrain}
\end{eqnarray}

Now, suppose $v_i$ as the source node posts a false message. The number of nodes in the network that post this false message can be estimated by
\begin{equation}
n_F(i)=\sum_{k=1}^{i-2}\left(\frac{\eta}{2}\right)^k+\sum_{k=0}^{N-i}\eta^k,
\label{Eq:CNFTrain}
\end{equation}
when $i\geq{2}$. Similarly, the average FTA of the network can be calculated by
\begin{eqnarray}
F_F&=&\frac{\sum_{i=2}^N{n_F(i)}}{N(N-1)}\nonumber\\
&=&\frac{1}{N(N-1)}\left[\frac{\eta}{2}\frac{N-1}{1-\eta/2}-\frac{\eta}{2}\frac{1-(\eta/2)^{N-1}}{(1-\eta/2)^2}\right]\nonumber\\
&+&\frac{1}{N(N-1)}\left[\frac{N-1}{1-\eta}-\eta\frac{1-\eta^{N-1}}{(1-\eta)^2}\right],
\label{Eq:FFChain1}
\end{eqnarray}
which can be simplified to
\begin{eqnarray}
F_F&=&\frac{1}{N}\frac{2+\eta-2\eta^2}{(2-\eta)(1-\eta)}\nonumber\\
&-&\frac{1}{N(N-1)}\left[\frac{2\eta}{(2-\eta)^2}+\frac{\eta}{(1-\eta)^2}\right],
\label{Eq:FFChainSTrain}
\end{eqnarray}
as $N\to\infty$. Based on Eqs.~(\ref{Eq:IFA}), (\ref{Eq:FTChainSTrain}) and (\ref{Eq:FFChainSTrain}), one can estimate the IFA by Eq.~(\ref{Eq:IFAChainTrain}).
\newcounter{mytempeqncnt}
\begin{figure*}[!t]
\normalsize
\begin{eqnarray}
\label{Eq:IFAChainTrain}
F&=&\frac{F_T-F_F}{F_F}\nonumber\\
&=&\frac{2\eta(1-\eta)^2+\eta(2-\eta)^2}{N\left[(2+\eta-2\eta^2)(2-\eta)(1-\eta)(N-1)-2\eta(1-\eta)^2-\eta(2-\eta)^2\right]}\sim\frac{2\eta(1-\eta)^2+\eta(2-\eta)^2}{N^2(2+\eta-2\eta^2)(2-\eta)(1-\eta)}.
\end{eqnarray}
\hrulefill
\vspace*{4pt}
\end{figure*}

By comparing Eq.~(\ref{Eq:IFAChain}) and Eq.~(\ref{Eq:IFAChainTrain}), it can be seen that, although the IFA of the re-weighted chain network will also diminish as the network size increases, it is always positive and indeed larger than the IFA of the original chain network. This completes the proof.

\subsection{Star Network}
\begin{figure}[!t]
\centering
\includegraphics[width=\linewidth]{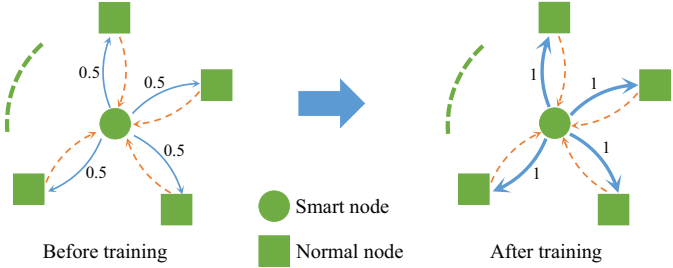}\caption{A star network with the center being the only smart node, before and after training.}
\label{StarNetwork}
\end{figure}

In this setting, assume that the center of the star network is a smart node, all the leafs are normal nodes, and all the weights of the directed links are set to 0.5 initially, as shown in Fig.~\ref{StarNetwork}. Since the center of the star network plays an important role in information diffusion, the following theorem is concerned with the center.

\emph{\textbf{Theorem 3:}} The IFA of a star network with the center being the only smart node increases as the NFR $\eta$ or the network size $N$ increases. And it is much larger than that of the chain network of the same size, suggesting that star networks are better modules for constructing networks with a higher information filtering ability.

\emph{\textbf{Proof:}} Without loss of generality, denote the only smart node as $v_1$. Then, all the normal nodes are connected to $v_1$, and there is no link between normal nodes. Now, suppose $v_i$ is the source node which posts a true message. The number of nodes in the network that consequently post the message can be estimated by
\begin{equation}
n_T(i)=1+\eta+\frac{\eta^2}{2}(N-2),
\label{Eq:SNT1}
\end{equation}
when $i\geq{2}$, and
\begin{equation}
n_T(i)=1+\frac{\eta}{2}(N-1),
\label{Eq:SNT2}
\end{equation}
when $i=1$. Suppose each node has an equal probability to be selected as the source node. Then, the average TTA of the network can be calculated by
\begin{eqnarray}
F_T&=&\frac{1}{N}\left[\frac{n_T(1)}{N}+\frac{\sum_{i=2}^N{n_T(i)}}{N}\right]\nonumber\\
&=&\frac{1}{N^2}\left[1+\frac{\eta}{2}(N-1)\right]\nonumber\\
&+&\frac{1}{N^2}\left[1+\eta+\frac{\eta^2}{2}(N-2)\right](N-1)\nonumber\\
&=&\frac{1}{N^2}\left[N+\frac{3\eta}{2}(N-1)+\frac{\eta^2}{2}(N^2-3N+2)\right]
\label{Eq:FTStar}
\end{eqnarray}

Suppose $v_i$ as the source node posts a false message. In this case, $v_i$ must be a normal node and it will be the only node that posts this message, since the smart node will never forward any false message, and thus all the other normal nodes cannot observe this message. Therefore, the number of nodes in the network that post the message is 1. And the average FTA of the star network can be calculated by
\begin{equation}
F_F=\frac{1}{N}.
\label{Eq:FFStar}
\end{equation}
Based on Eqs.~(\ref{Eq:IFA}), (\ref{Eq:FTStar}) and (\ref{Eq:FFStar}), one can calculate the IFA as follows:
\begin{eqnarray}
F&=&\frac{F_T-F_F}{F_F}\nonumber\\
&=&\frac{1}{N}\left[\frac{3\eta}{2}(N-1)+\frac{\eta^2}{2}(N^2-3N+2)\right].
\label{Eq:IFAStar}
\end{eqnarray}

Here, one can see that $F$ is an increasing function of $\eta$ and $N$, indicating that the IFA of a star network can be enhanced by increasing the NFR of the nodes or the network size. By comparing  Eq.~(\ref{Eq:IFAStar}) and Eq.~(\ref{Eq:IFAChain}), one can see that the IFA of a star network is indeed much larger than that of the chain network of the same size. This completes the proof.

\emph{\textbf{Remark 2:}} When a star network with the center being the only smart node is trained, the weights of all the directed links from the smart node to normal nodes increase with time, and will become 1 finally, since  the normal nodes always observe true message from the smart node. The weights of the directed links from normal nodes to the smart node are useless, since the smart node makes decision independently to forward a message or not, i.e., it will forward the true message with probability $\eta$ but never forward a false message.

\emph{\textbf{Theorem 4:}} After training, the IFA of the star network with the center being a smart node becomes larger, indicating that the self-learning mechanism enhances the star network to become smarter, in the sense of distinguishing a true message from the false.

\emph{\textbf{Proof:}} After training, suppose $v_i$ is the source node which posts a true message. The number of nodes in the network that consequently post this true message can be estimated by
\begin{equation}
n_T(i)=1+\eta+\eta^2(N-2),
\label{Eq:SNTTrain1}
\end{equation}
when $i\geq{2}$, and
\begin{equation}
n_T(i)=1+\eta(N-1),
\label{Eq:SNTTrain2}
\end{equation}
when $i=1$. Suppose each node has an equal probability to be selected as the source node. Then, the average TTA of the network can be calculated by
\begin{eqnarray}
F_T&=&\frac{1}{N}\left[\frac{n_T(1)}{N}+\frac{\sum_{i=2}^N{n_T(i)}}{N}\right]\nonumber\\
&=&\frac{1}{N^2}\left[1+\eta(N-1)\right]\nonumber\\
&+&\frac{N-1}{N^2}\left[1+\eta+\eta^2(N-2)\right]\nonumber\\
&=&\frac{1}{N^2}\left[N+2\eta(N-1)+\eta^2(N^2-3N+2)\right].
\label{Eq:FTStarTrain}
\end{eqnarray}

For the false message, one obtains a completely same result as that on the untrained network, because it will never be posted by the smart node so that all the normal nodes except the source node cannot observe this message. Thus, the average FTA of the star network can be calculated by
\begin{equation}
F_F=\frac{1}{N}.
\label{Eq:FFStarTrain}
\end{equation}
Based on Eqs.~(\ref{Eq:IFA}), (\ref{Eq:FTStarTrain}) and (\ref{Eq:FFStarTrain}), one can calculate the IFA as follows:
\begin{eqnarray}
F&=&\frac{F_T-F_F}{F_F}\nonumber\\
&=&\frac{1}{N}\left[2\eta(N-1)+\eta^2(N^2-3N+2)\right].
\label{Eq:IFAStarTrain}
\end{eqnarray}

By comparing Eq.~(\ref{Eq:IFAStarTrain}) and Eq.~(\ref{Eq:IFAStar}), one can see that the IFA of the star network is almost doubled after training, which indicates that star networks have higher potential than chain networks to become smarter, in the sense of better distinguishing a true message from the false, by adopting the self-learning mechanism. This completes the proof.

\subsection{Emergence of Social Stratification}
Generally, \emph{social stratification} is a relative social position of an individual within a social group, mainly based on his occupation and income, wealth and social status, or derived power. In this study, however, we define the social stratification of a node $v_i$ purely based on the information flows, i.e., the number of nodes that forward the true or false message initially posted by node $v_i$.  This can be considered as a certain power of information diffusion, which is quite important in the information era today.

\emph{\textbf{Lemma 1:}} For the chain network with one terminal being a smart node, social stratification based on information diffusion emerges due to the introduction of the smart node, which is further enhanced by the self-learning mechanism.

\emph{\textbf{Proof:}} In particular, consider the difference of information diffusion power between two successive normal nodes $v_{i}$ and $v_{i+1}$ in a chain network. Before training, for the true massage, based on Eq.~(\ref{Eq:CNT1}), one has
\begin{eqnarray}
D_T(i)&=&n_T(i)-n_T(i+1)\nonumber\\
&=&\left(\frac{\eta}{2}\right)^{N-i}-\left(\frac{\eta}{2}\right)^{i-1}+\eta\left(\frac{\eta}{2}\right)^{i-2}\left(1-\frac{\eta}{2}\right)\nonumber\\
&=&\left(\frac{\eta}{2}\right)^{N-i}+(1-\eta)\left(\frac{\eta}{2}\right)^{i-1}.
\label{Eq:DTrue}
\end{eqnarray}
For the false message, based on Eq~(\ref{Eq:CNF}), one has
\begin{eqnarray}
D_F(i)&=&n_F(i)-n_F(i+1)\nonumber\\
&=&\left(\frac{\eta}{2}\right)^{N-i}-\left(\frac{\eta}{2}\right)^{i-1}.
\label{Eq:DFalse}
\end{eqnarray}

\emph{\textbf{Remark 3:}} From Eq.~(\ref{Eq:DTrue}), one always has $D_T(i)>0$ when $0<\eta\leq{1}$, indicating a social stratification, from the smart node to the other terminal, that the normal nodes closer to the smart node have higher powers to deliver true messages, i.e., their true messages could be forwarded by more other nodes. From Eq.~(\ref{Eq:DFalse}), on the other hand, one has $D_F(i)>0$ when $i>(N+1)/2$ and $D_F(i)<0$ when $i<(N+1)/2$, indicating that the normal node $v_m$ in the middle of the chain, with $m=\lceil(N+1)/2\rceil$, has the highest power to deliver a false message, and this power decreases steadily from the middle to the terminals, less influenced by the smart node.

After training, for the true massage, based on Eq.~(\ref{Eq:CNTTrain1}), the difference of information diffusion power between two successive normal nodes $v_{i}$ and $v_{i+1}$ is calculated by
\begin{eqnarray}
D_T(i)&=&n_T(i)-n_T(i+1)\nonumber\\
&=&\eta^{N-i}-\left(\frac{\eta}{2}\right)^{i-1}+\eta\left(\frac{\eta}{2}\right)^{i-2}\left(1-\frac{\eta}{2}\right)\nonumber\\
&=&\eta^{N-i}+(1-\eta)\left(\frac{\eta}{2}\right)^{i-1}.
\label{Eq:DTrueTrain}
\end{eqnarray}
For the false message, based on Eq~(\ref{Eq:CNFTrain}), it is calculated by
\begin{eqnarray}
D_F(i)&=&n_F(i)-n_F(i+1)\nonumber\\
&=&\eta^{N-i}-\left(\frac{\eta}{2}\right)^{i-1}.
\label{Eq:DFalseTrain}
\end{eqnarray}

\emph{\textbf{Remark 4:}} Similarly, Eq.~(\ref{Eq:DTrueTrain}) tells that one will always have $D_T(i)>0$ when $0<\eta\leq{1}$, indicating the same social stratification from the smart node to the other terminal, in the sense of decreasing the information diffusion power of true messages. By comparing Eq.~(\ref{Eq:DTrueTrain}) and Eq.~(\ref{Eq:DTrue}), one can see that $D_T(i)$ has a larger value in the trained network than in the original network. For Eq.~(\ref{Eq:DFalseTrain}), letting $D_F(i)=0$ gives
\begin{eqnarray}
&&\eta^{N-i}=\left(\frac{\eta}{2}\right)^{i-1}\nonumber\\
&\Rightarrow&(N-i)\ln\eta=(i-1)\ln\frac{\eta}{2}\nonumber\\
&\Rightarrow&\left(\ln\eta+\ln\frac{\eta}{2}\right)i=N\ln\eta+\ln\frac{\eta}{2}\nonumber\\
&\Rightarrow&i=\frac{(N+1)\ln\eta-\ln2}{2\ln\eta-\ln2}<\frac{N+1}{2}.
\label{Eq:HPower}
\end{eqnarray}
Eq.~(\ref{Eq:HPower}) indicates that the normal node of the highest power to deliver the false messages move towards the smart node, and such tendency is more prominent for larger values of $\eta$. In particular, when $\eta=1$, one will always have $D_F(i)>0$ for $i\geq{2}$, showing the similar social stratification from the smart node to the other terminal as for the case of true messages.

These results suggest that the social stratification introduced by the smart node is enhanced by the self-learning mechanism. This completes the proof.

\subsection{Crossover Advantage\label{CA}}
\begin{figure}[!ht]
\centering
\includegraphics[width=\linewidth]{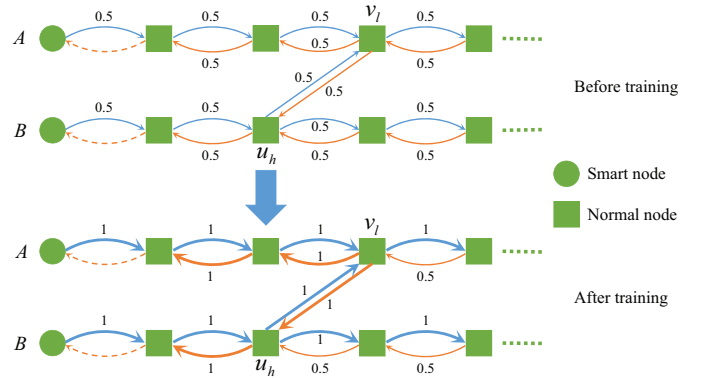}\caption{The interconnection between two chain networks with one terminal being the smart node, before and after training.}
\label{CrossBorder}
\end{figure}

Now, consider two chain networks, $A$ and $B$, each has $N$ nodes with one terminal being the smart node.  If there is no interconnection between the two networks, the analysis will be the same as that in Sec.~\ref{chain}. Now, add an interconnection between two nodes from $A$ and $B$, denoted by $v_l$ and $u_h$, respectively, as shown in Fig.~\ref{CrossBorder}.

\emph{\textbf{Lemma 2:}} The interconnection between two different chain networks will increase the social influences of the bridge nodes, i.e., $v_l$ and $u_h$ in Fig.~\ref{CrossBorder}, reflecting the \emph{crossover advantage}. And such advantage could be enlarged due to the self-learning mechanism.

\emph{\textbf{Proof:}} Suppose $v_i$ in $A$ is the source node which posts a true message. The number of nodes in the whole interconnected network that consequently post the message can be estimated by Eq.~(\ref{Eq:CBT1}) when $i\geq{2}$,
\begin{figure*}[!t]
\normalsize
\begin{eqnarray}
\label{Eq:CBT1}
n_T^A(i) &=&\sum_{k=0}^{N-i}\left(\frac{\eta}{2}\right)^k+\sum_{k=1}^{i-2}\left(\frac{\eta}{2}\right)^k+\eta\left(\frac{\eta}{2}\right)^{i-2}
+\left(\frac{\eta}{2}\right)^{|i-l|+1}\left[\sum_{k=0}^{N-h}\left(\frac{\eta}{2}\right)^k+\sum_{k=1}^{h-2}\left(\frac{\eta}{2}\right)^k+\eta\left(\frac{\eta}{2}\right)^{h-2}\right]
\end{eqnarray}
\hrulefill
\vspace*{4pt}
\end{figure*}
 and by
\begin{eqnarray}
n_T^A(i)&=&\sum_{k=0}^{N-1}\left(\frac{\eta}{2}\right)^k\nonumber\\
&+&\left(\frac{\eta}{2}\right)^l\left[\sum_{k=0}^{N-h}\left(\frac{\eta}{2}\right)^k+\sum_{k=1}^{h-2}\left(\frac{\eta}{2}\right)^k+\eta\left(\frac{\eta}{2}\right)^{h-2}\right],\nonumber\\
\label{Eq:CBT2}
\end{eqnarray}
when $i=1$. Suppose $u_i$ in $B$ is the source node, which posts a true message. One can exchange the places of $h$ and $l$ in Eqs.~(\ref{Eq:CBT1}) and (\ref{Eq:CBT2}) to get the corresponding $n_T^B(i)$ for $i\geq{2}$ and $i=1$, respectively.

Suppose $v_i$ as the source node posts a false message. The number of nodes in the whole network that consequently post this false message can be estimated by
\begin{eqnarray}
n_F^A(i)&=&\sum_{k=1}^{i-2}\left(\frac{\eta}{2}\right)^k+\sum_{k=0}^{N-i}\left(\frac{\eta}{2}\right)^k\nonumber\\
&+&\left(\frac{\eta}{2}\right)^{|i-l|+1}\left[\sum_{k=1}^{h-2}\left(\frac{\eta}{2}\right)^k+\sum_{k=0}^{N-h}\left(\frac{\eta}{2}\right)^k\right],
\label{Eq:CBF}
\end{eqnarray}
when $i\geq{2}$. Similarly, by exchanging the places of $h$ and $l$ in Eq.~(\ref{Eq:CBF}), one can get the corresponding $n_F^B(i)$.

Now, consider how the interconnection changes the social stratification in a chain network. Take the chain network $A$ for example, and consider the difference of information diffusion power between two successive normal nodes $v_{i}$ and $v_{i+1}$. Before training, for the true massage, based on Eq.~(\ref{Eq:CBT1}), one has
\begin{eqnarray}
D_T^A(i)&=&n_T^A(i)-n_T^A(i+1)\nonumber\\
&=&\left(\frac{\eta}{2}\right)^{N-i}+(1-\eta)\left(\frac{\eta}{2}\right)^{i-1}\nonumber\\
&+&\left[\left(\frac{\eta}{2}\right)^{|i-l|+1}-\left(\frac{\eta}{2}\right)^{|i+1-l|+1}\right]\theta_T,
\label{Eq:CBDTrue}
\end{eqnarray}
where $\theta_T$ is defined by
\begin{equation}
\theta_T=\sum_{k=0}^{N-h}\left(\frac{\eta}{2}\right)^k+\sum_{k=1}^{h-2}\left(\frac{\eta}{2}\right)^k+\eta\left(\frac{\eta}{2}\right)^{h-2},
\label{Eq:Theta}
\end{equation}
which is independent of $i$ and thus can be considered as a positive constant. From Eq.~(\ref{Eq:CBDTrue}), one can easily obtain
\begin{equation}
D_T^A(i)<\left(\frac{\eta}{2}\right)^{N-i}+(1-\eta)\left(\frac{\eta}{2}\right)^{i-1},
\label{Eq:CBDTrue1}
\end{equation}
when $i<l$, and
\begin{equation}
D_T^A(i)>\left(\frac{\eta}{2}\right)^{N-i}+(1-\eta)\left(\frac{\eta}{2}\right)^{i-1},
\label{Eq:CBDTrue2}
\end{equation}
when $i\geq{l}$. By comparing Eqs.~(\ref{Eq:CBDTrue1}), (\ref{Eq:CBDTrue2}), and~(\ref{Eq:DTrue}), one may conclude that, before training and for the true message, the social stratification between the nodes in the sub-chain from $v_1$ to $v_l$ (or from $u_1$ to $u_h$) is weakened or even reversed, while that between the nodes in the rest sub-train from $v_l$ to $v_N$ (or from $u_h$ to $u_N$) is strengthened, with an interconnection between node $v_l$ (or $u_h$) and any normal node in the other chain network.

For the false message, based on Eq.~(\ref{Eq:CBF}), before training, the difference of information diffusion power between two successive normal nodes $v_{i}$ and $v_{i+1}$ is calculated by
\begin{eqnarray}
D_F^A(i)&=&n_F^A(i)-n_F^A(i+1)\nonumber\\
&=&\left(\frac{\eta}{2}\right)^{N-i}-\left(\frac{\eta}{2}\right)^{i-1}\nonumber\\
&+&\left[\left(\frac{\eta}{2}\right)^{|i-l|+1}-\left(\frac{\eta}{2}\right)^{|i+1-l|+1}\right]\theta_F,
\label{Eq:CBDFalse}
\end{eqnarray}
where $\theta_F$ is defined by
\begin{equation}
\theta_F=\sum_{k=0}^{N-h}\left(\frac{\eta}{2}\right)^k+\sum_{k=1}^{h-2}\left(\frac{\eta}{2}\right)^k,
\label{Eq:ThetaF}
\end{equation}
which is independent of $i$ and thus can be considered as a positive constant. In this case, by comparing Eqs.~(\ref{Eq:CBDFalse}) and (\ref{Eq:DFalse}), one can see that, before training and for the false message, the social stratification between the nodes $v_i$ with $i<\min\{l,(N+1)/2\}$ or $i>\max\{l,(N+1)/2\}$ is strengthened, while that between the other nodes is weakened or even reversed. When considering the nodes in the chain network $B$, the results are similar, obtained via replacing $v_i$ by $u_i$ and $l$ by $h$.

The above results suggest that the interconnection between different chain networks will increase the social influences of the bridge nodes, i.e., $v_l$ and $u_h$ in Fig.~\ref{CrossBorder}, indicating the crossover advantage.

\emph{\textbf{Remark 5:}} In the following, it is to study whether there is still such crossover advantage after the training process. For a single chain network, it has been proved that, after training, the weights of the directed links from $v_{i-1}$ to $v_i$ tend to be 1, while those from $v_{i}$ to $v_{i-1}$ will be close to 0.5,  for $i=2,3,\ldots,N$, as shown in Fig.~\ref{ChainNetwork}. For the same reasons, for the chain network $A$ (or $B$) here, after training, the weights of the directed links from $v_{i-1}$ to $v_i$ (or from $u_{i-1}$ to $u_i$) tend to be 1, for $i=2,3,\ldots,N$. However, due to the interconnection between $A$ and $B$, the true message posted by $v_1$ could be delivered to $u_h$, and further to $u_{h-k}$, for $k=1,2,\ldots,h-2$. This will make the weights of the directed links from $v_l$ to $u_h$ and also $u_{i}$ to $u_{i-1}$, for $i=3,\ldots,h$,  tend to be 1. Correspondingly, the weights of the directed links from $u_h$ to $v_l$ and $v_{i}$ to $v_{i-1}$, for $i=3,\ldots,l$, also tend to be 1. And the rest links will have weights close to 0.5, as shown in Fig.~\ref{CrossBorder}.

After training, suppose $v_i$ in $A$ is the source node which posts a true message. The number of nodes in the whole network that consequently post the message can be estimated by
\begin{eqnarray}
n_T^A(i) &=&\sum_{k=0}^{N-i}\eta^k+\sum_{k=1}^{i-1}\eta^k \nonumber\\
&+&\eta^{l-i+1}\left[\sum_{k=0}^{N-h}\eta^k+\sum_{k=1}^{h-1}\eta^k\right],
\label{Eq:CBTTrain1}
\end{eqnarray}
when $2\leq{i}\leq{l}$; by
\begin{eqnarray}
n_T^A(i) &=&\sum_{k=0}^{N-i}\eta^k+\left(\frac{\eta}{2}\right)^{i-l}\sum_{k=1}^{l-1}\eta^k+\sum_{k=1}^{i-l}\left(\frac{\eta}{2}\right)^{k} \nonumber\\
&+&\eta\left(\frac{\eta}{2}\right)^{i-l}\left[\sum_{k=0}^{N-h}\eta^k+\sum_{k=1}^{h-1}\eta^k\right],
\label{Eq:CBTTrain2}
\end{eqnarray}
when $l<{i}\leq{N}$; and by
\begin{eqnarray}
n_T^A(i)&=&\sum_{k=0}^{N-1}\eta^k+\eta^l\left[\sum_{k=0}^{N-h}\eta^k+\sum_{k=1}^{h-1}\eta^k\right],
\label{Eq:CBTTrain3}
\end{eqnarray}
when $i=1$. Suppose $u_i$ in $B$ is the source node, which posts a true message. One can simply exchange the places of $h$ and $l$ in Eqs.~(\ref{Eq:CBTTrain1})-(\ref{Eq:CBTTrain3}) to get the corresponding $n_T^B(i)$ in different situations.

Suppose $v_i$ in $A$, as the source node, posts a false message. Then, the number of nodes in the whole network that consequently post the false message can be estimated by
\begin{eqnarray}
n_F^A(i) &=&\sum_{k=0}^{N-i}\eta^k+\sum_{k=1}^{i-2}\eta^k \nonumber\\
&+&\eta^{l-i+1}\left[\sum_{k=0}^{N-h}\eta^k+\sum_{k=1}^{h-2}\eta^k\right],
\label{Eq:CBFTrain1}
\end{eqnarray}
when $2\leq{i}\leq{l}$ and by
\begin{eqnarray}
n_F^A(i) &=&\sum_{k=0}^{N-i}\eta^k+\left(\frac{\eta}{2}\right)^{i-l}\sum_{k=1}^{l-2}\eta^k+\sum_{k=1}^{i-l}\left(\frac{\eta}{2}\right)^{k} \nonumber\\
&+&\eta\left(\frac{\eta}{2}\right)^{i-l}\left[\sum_{k=0}^{N-h}\eta^k+\sum_{k=1}^{h-2}\eta^k\right],
\label{Eq:CBFTrain2}
\end{eqnarray}
when $l<{i}<{N}$. Similarly, by exchanging the places of $h$ and $l$ in Eqs.~(\ref{Eq:CBFTrain1}) and (\ref{Eq:CBFTrain2}), one can get the corresponding $n_F^B(i)$ in different situations.

This time, one has
\begin{eqnarray}
D_T^A(i)&=&n_T^A(i)-n_T^A(i+1)\nonumber\\
&=&\eta^{N-i}-\eta^i-\eta^{l-i}(1-\eta)\beta_T\nonumber\\
&<&\eta^{N-i}<\eta^{N-i}+(1-\eta)\left(\frac{\eta}{2}\right)^{i-1},
\label{Eq:CBDTrueTrain1}
\end{eqnarray}
when $2\leq{i}\leq{l-1}$, and
\begin{eqnarray}
D_T^A(i)&=&\eta^{N-i}+\left(\frac{\eta}{2}\right)^{i-l}\left(1-\frac{\eta}{2}\right)\sum_{k=1}^{l-1}\eta^k\nonumber\\
&-&\left(\frac{\eta}{2}\right)^{i-l+1}+\eta\left(1-\frac{\eta}{2}\right)\left(\frac{\eta}{2}\right)^{i-l}\beta_T,
\label{Eq:CBDTrueTrain2}
\end{eqnarray}
when $l\leq{i}<N$, where $\beta_T$ is defined by
\begin{equation}
\beta_T=\sum_{k=0}^{N-h}\eta^k+\sum_{k=1}^{h-1}\eta^k,
\label{Eq:BetaT}
\end{equation}
which is independent of $i$ and thus can be considered as a positive constant. From Eq.~(\ref{Eq:BetaT}), one can easily verify that $\beta_T>1$. Since $0<\eta\leq{1}$, one obtains that
\begin{eqnarray}
\eta\left(1-\frac{\eta}{2}\right)\left(\frac{\eta}{2}\right)^{i-l}\beta_T&>&\eta\left(1-\frac{\eta}{2}\right)\left(\frac{\eta}{2}\right)^{i-l}\nonumber\\
&=&(2-\eta)\left(\frac{\eta}{2}\right)^{i-l+1}\nonumber\\
&\geq&\left(\frac{\eta}{2}\right)^{i-l+1}.
\label{Ineq1}
\end{eqnarray}
For $l\geq{2}$, one has $\sum_{k=1}^{l-1}\eta^k\geq{\eta}$, so that based on Eq.~(\ref{Ineq1}), Eq.~(\ref{Eq:CBDTrueTrain2}) is changed to
\begin{eqnarray}
D_T^A(i)&>&\eta^{N-i}+\left(\frac{\eta}{2}\right)^{i-l}\left(1-\frac{\eta}{2}\right)\sum_{k=1}^{l-1}\eta^k\nonumber\\
&\geq&\eta^{N-i}+\left(\frac{\eta}{2}\right)^{i-l}\left(1-\frac{\eta}{2}\right)\eta\nonumber\\
&\geq&\eta^{N-i}+\left(\frac{\eta}{2}\right)^{i-l+1}(1-\eta)\nonumber\\
&\geq&\eta^{N-i}+\left(\frac{\eta}{2}\right)^{i-1}(1-\eta).
\label{Ineq2}
\end{eqnarray}

For the false message, based on Eqs.~(\ref{Eq:CBFTrain1}) and (\ref{Eq:CBFTrain2}), after training, the difference of information diffusion power between two successive normal nodes $v_{i}$ and $v_{i+1}$ is calculated by
\begin{eqnarray}
D_F^A(i)&=&n_F^A(i)-n_F^A(i+1)\nonumber\\
&=&\eta^{N-i}-\eta^{i-1}-\eta^{l-i}(1-\eta)\beta_F\nonumber\\
&\leq&\eta^{N-i}-\eta^{i-1}\nonumber\\
&\leq&\eta^{N-i}-\left(\frac{\eta}{2}\right)^{i-1},
\label{Eq:CBDFalseTrain1}
\end{eqnarray}
when $2\leq{i}\leq{l}$, and by
\begin{eqnarray}
D_F^A(i)&=&\eta^{N-i}+\left(\frac{\eta}{2}\right)^{i-l}\left(1-\frac{\eta}{2}\right)\sum_{k=1}^{l-2}\eta^k\nonumber\\
&-&\left(\frac{\eta}{2}\right)^{i-l+1}+\eta\left(1-\frac{\eta}{2}\right)\left(\frac{\eta}{2}\right)^{i-l}\beta_F,
\label{Eq:CBDFalseTrain2}
\end{eqnarray}
when $l\leq{i}<N$, where $\beta_F$ is defined by
\begin{equation}
\beta_F=\sum_{k=0}^{N-h}\eta^k+\sum_{k=1}^{h-2}\eta^k,
\label{Eq:BetaF}
\end{equation}
which is independent of $i$ and thus can be considered as a positive constant. Therefore, Eq.~(\ref{Eq:CBDFalseTrain2}) is changed to
\begin{eqnarray}
D_F^A(i)&\geq&\eta^{N-i}-\left(\frac{\eta}{2}\right)^{i-l+1}\nonumber\\
&\geq&\eta^{N-i}-\left(\frac{\eta}{2}\right)^{i-1}.
\label{Ineq3}
\end{eqnarray}
These results still hold for the case where node $u_i$ in $B$ is chosen as the source node which posts the true message.

\begin{figure*}[!t]
\centering
\subfigure[Before training]{
\includegraphics[width=.28\linewidth]{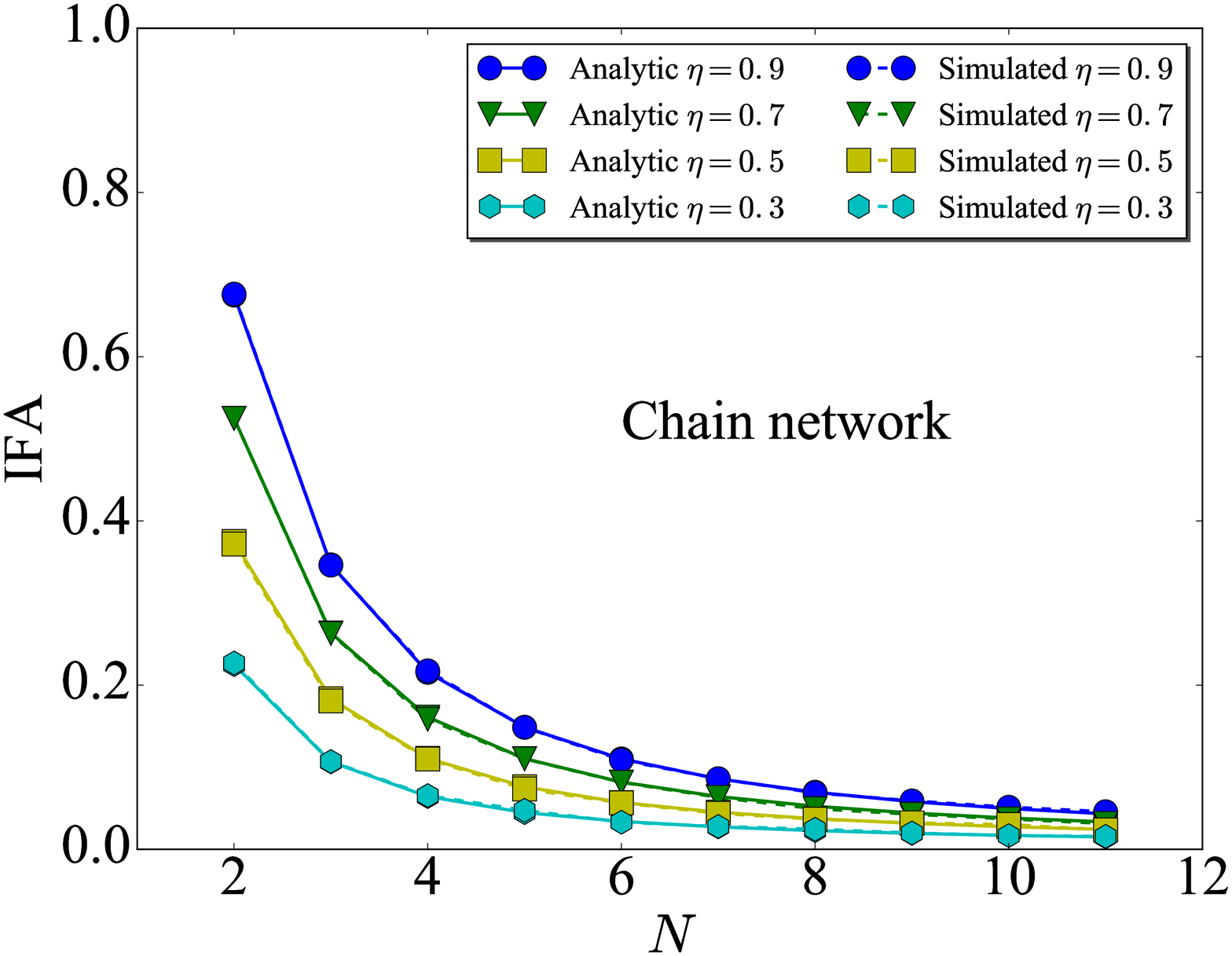}}
\subfigure[After training]{
\includegraphics[width=.28\linewidth]{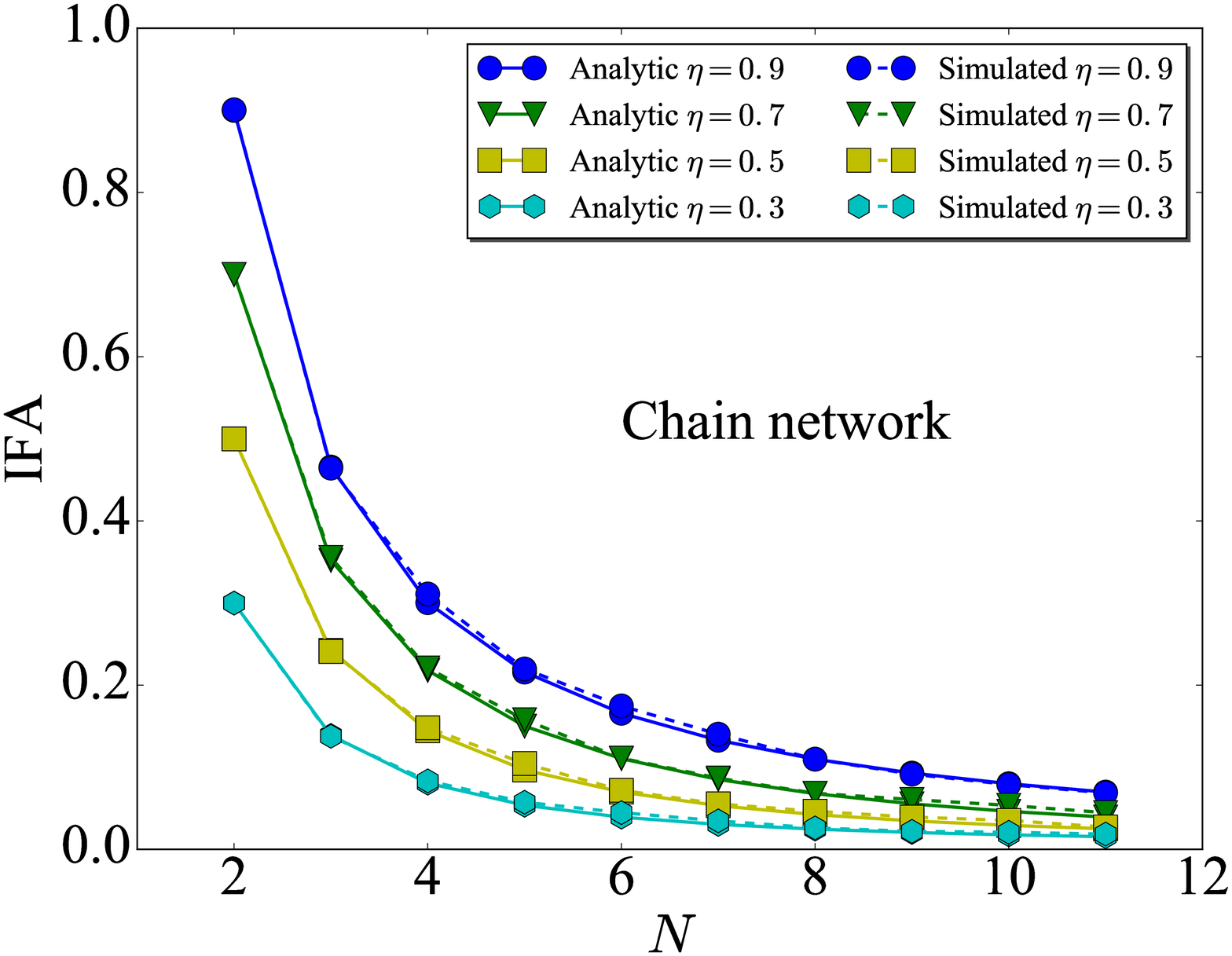}}
\subfigure[Relative improvement of IFA]{
\includegraphics[width=.40\linewidth]{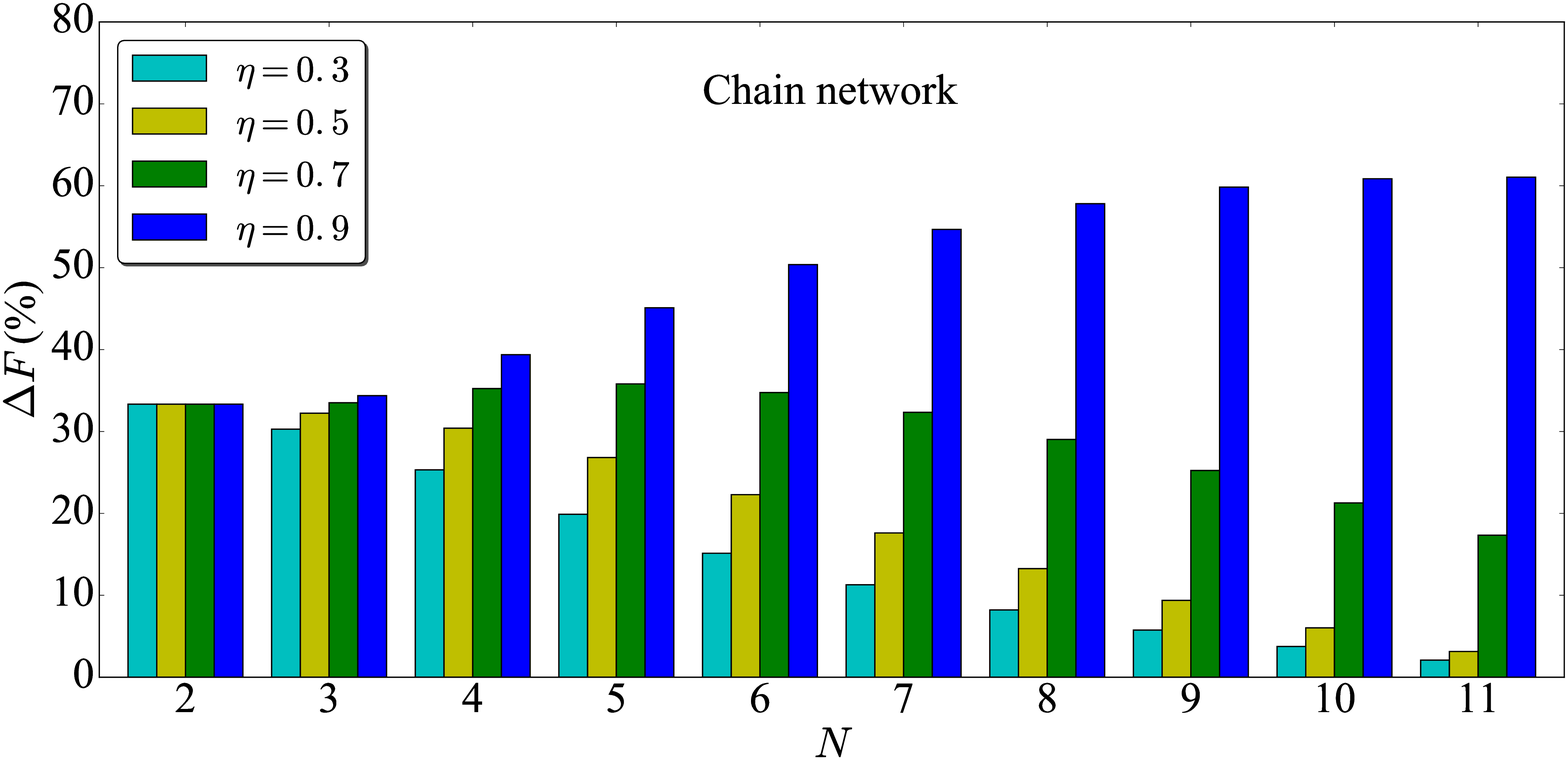}}
\caption{The analytic and simulated values of IFA as functions of the network size $N$ with various values of NFR $\eta$, for the chain network (a) before training and (b) after training. (c) The relative improvement of IFA by the training process.}
\label{exp1chain}
\end{figure*}

\begin{figure*}[!t]
\centering
\subfigure[Before training]{
\includegraphics[width=.28\linewidth]{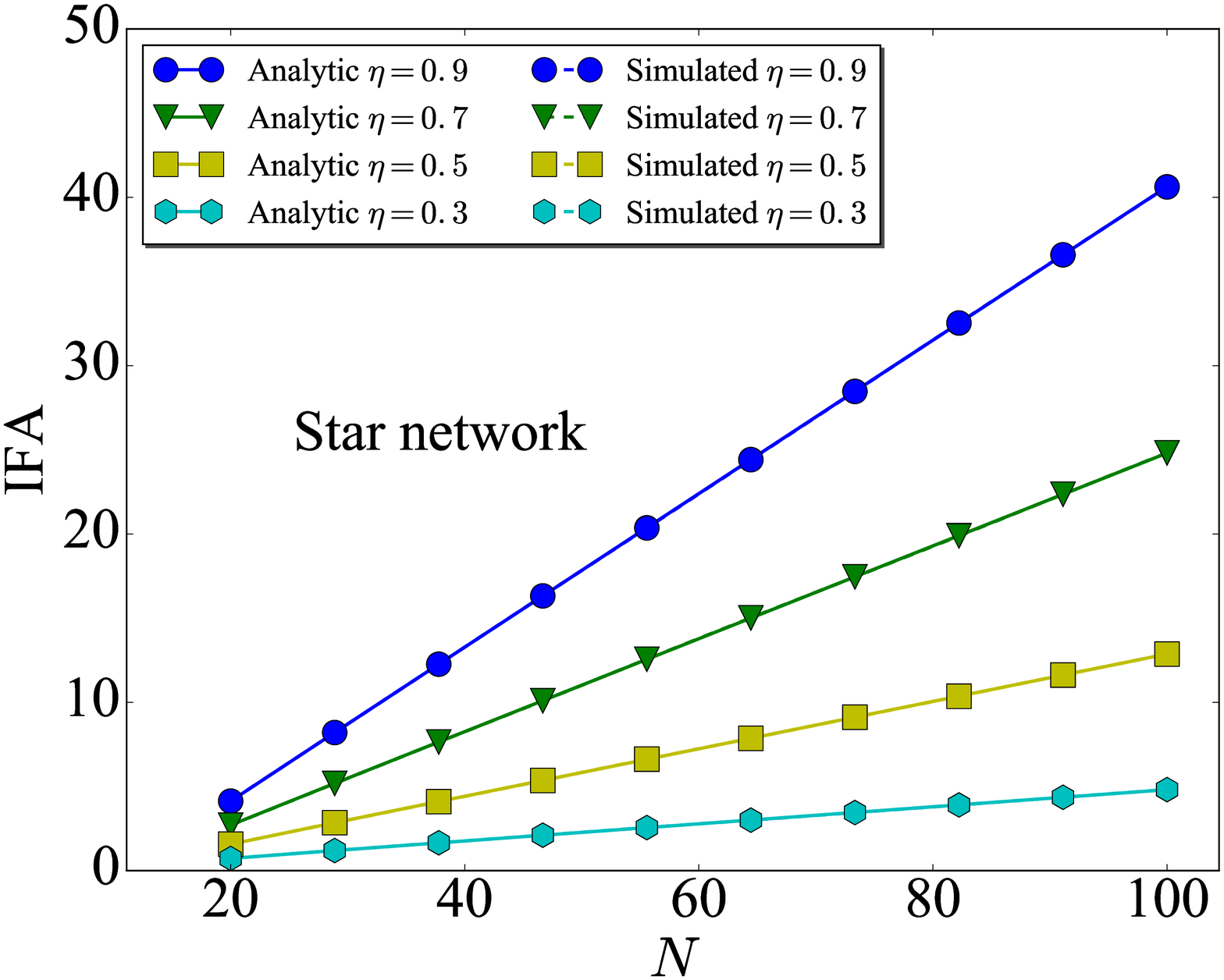}}
\subfigure[After training]{
\includegraphics[width=.28\linewidth]{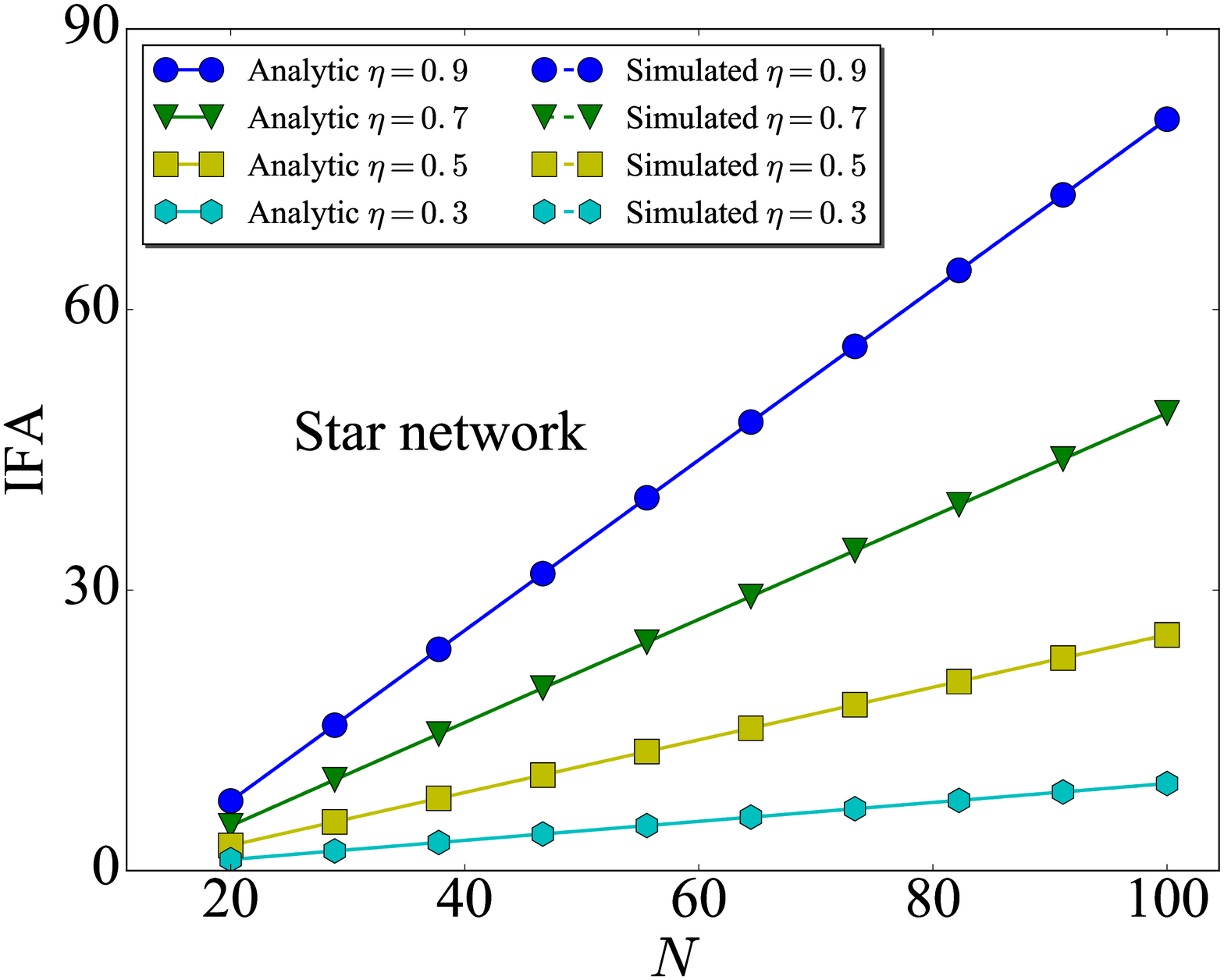}}
\subfigure[Relative improvement of IFA]{
\includegraphics[width=.40\linewidth]{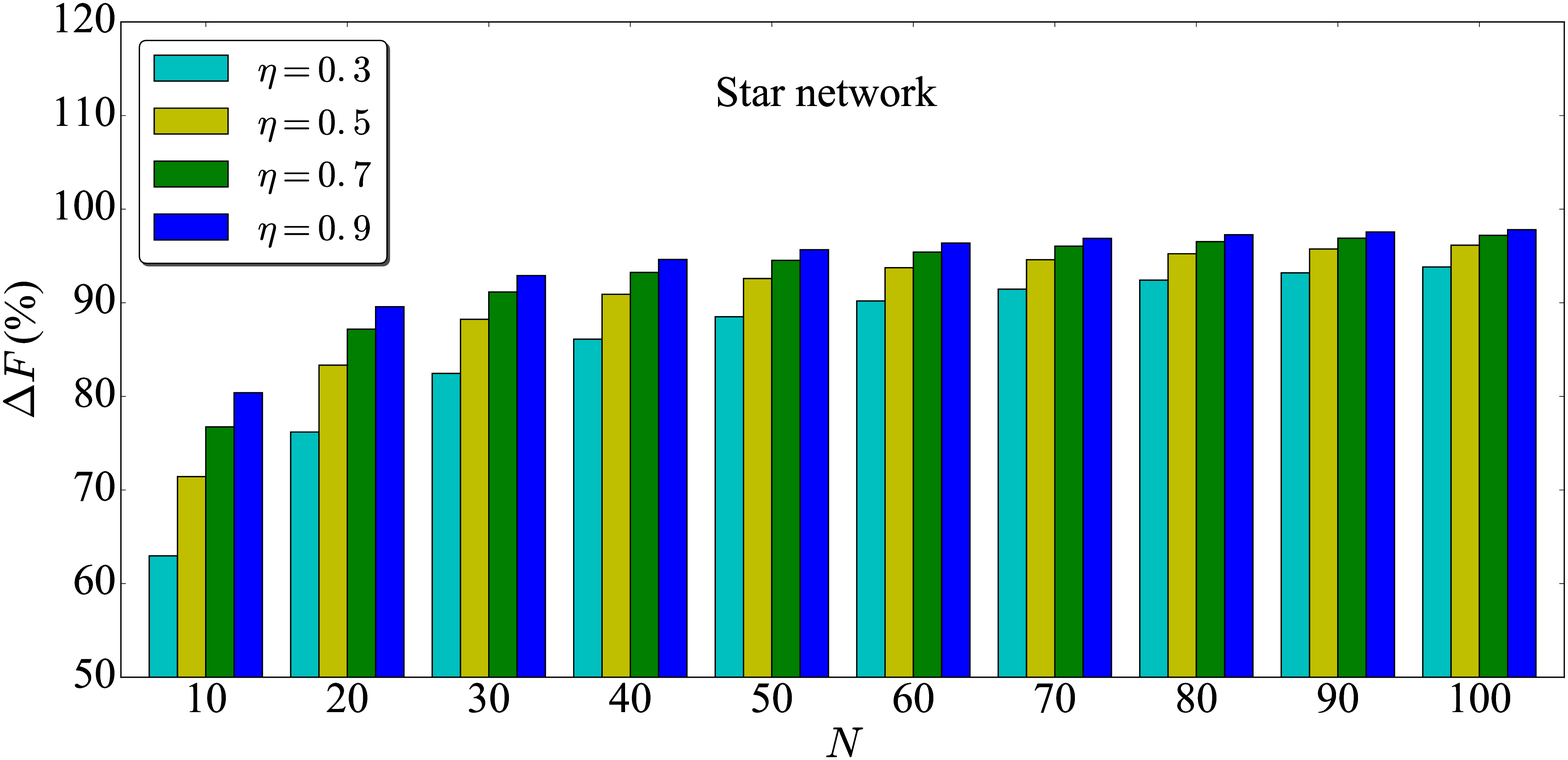}}
\caption{The analytic and simulated values of IFA as functions of the network size $N$ with various values of NFR $\eta$, for the star network (a) before training and (b) after training. (c) The relative improvement of IFA by the training process.}
\label{exp1star}
\end{figure*}

By comparing Eqs.~(\ref{Eq:CBDTrueTrain1}), (\ref{Ineq2}) and Eq.~(\ref{Eq:DTrueTrain}), and comparing Eqs.~(\ref{Eq:CBDFalseTrain1}), (\ref{Ineq3}) and Eq.~(\ref{Eq:DFalseTrain}),  one can also find that, after training, for both true and false messages, the social stratification between the nodes in the sub-chain from $v_1$ to $v_l$ (or from $u_1$ to $u_h$) is weakened or even reversed, while that between the nodes in the rest sub-train from $v_l$ to $v_N$ (or from $u_h$ to $u_N$) is strengthened, with an interconnection between node $v_l$ (or $u_h$) and any normal node in the other chain network. These suggest that the crossover advantage can be enlarged by the self-learning mechanism. This completes the proof.

\section{Numerical Results\label{NR}}
Now, the above analytic results are verified by simulations.

\subsection{Information Filtering Ability}
Eqs.~(\ref{Eq:IFAChain}), (\ref{Eq:IFAChainTrain}) and Eqs.~(\ref{Eq:IFAStar}), (\ref{Eq:IFAStarTrain}) indicate that the IFA of a chain network decreases, while that of a star network increases, as the network size increases, whether or not the network has been trained. In order to investigate such trends, consider a cascading model on two types of networks of various sizes, and then calculate their corresponding values of IFA.

Specifically, in simulations the size of the chain network is varied from 2 to 10, while the size of the star network is varied from 10 to 100. Here, consider only a smaller chain network, since the cascading is typically determined by the diameter of a network, and thus the chain network (with the diameter close to the network size) is far more difficult to train than the star network (with the diameter equal to 2, which is independent of the network size). In the simulations, 10,000 messages are sent out from each node and then the means of TTA and FTA are calculated, followed by IFA. For the training process, set $\delta=\Delta=0.001$ and iterate 4,000,000 times for the relatively small chain network and 2,000 times for the relatively large star network. It is found that, indeed, the IFA of the chain network decreases very fast, following $F\sim{N^{-2}}$, while that of the star network increases linearly, following $F\sim{N}$, as the network size increases, whether or not the network is trained. The analytic and simulated results match quite well, as shown in Figs.~\ref{exp1chain} (a), (b) and Figs.~\ref{exp1star} (a), (b), respectively. Besides, it is found that the values of IFA increase as the NFR $\eta$ increases, for both chain and star networks, whether or not the networks are trained, indicating that the networks may become smarter in the environment where the information can easily spread.

\begin{figure*}[!t]
\centering
\subfigure[Before training]{
\includegraphics[width=.28\linewidth]{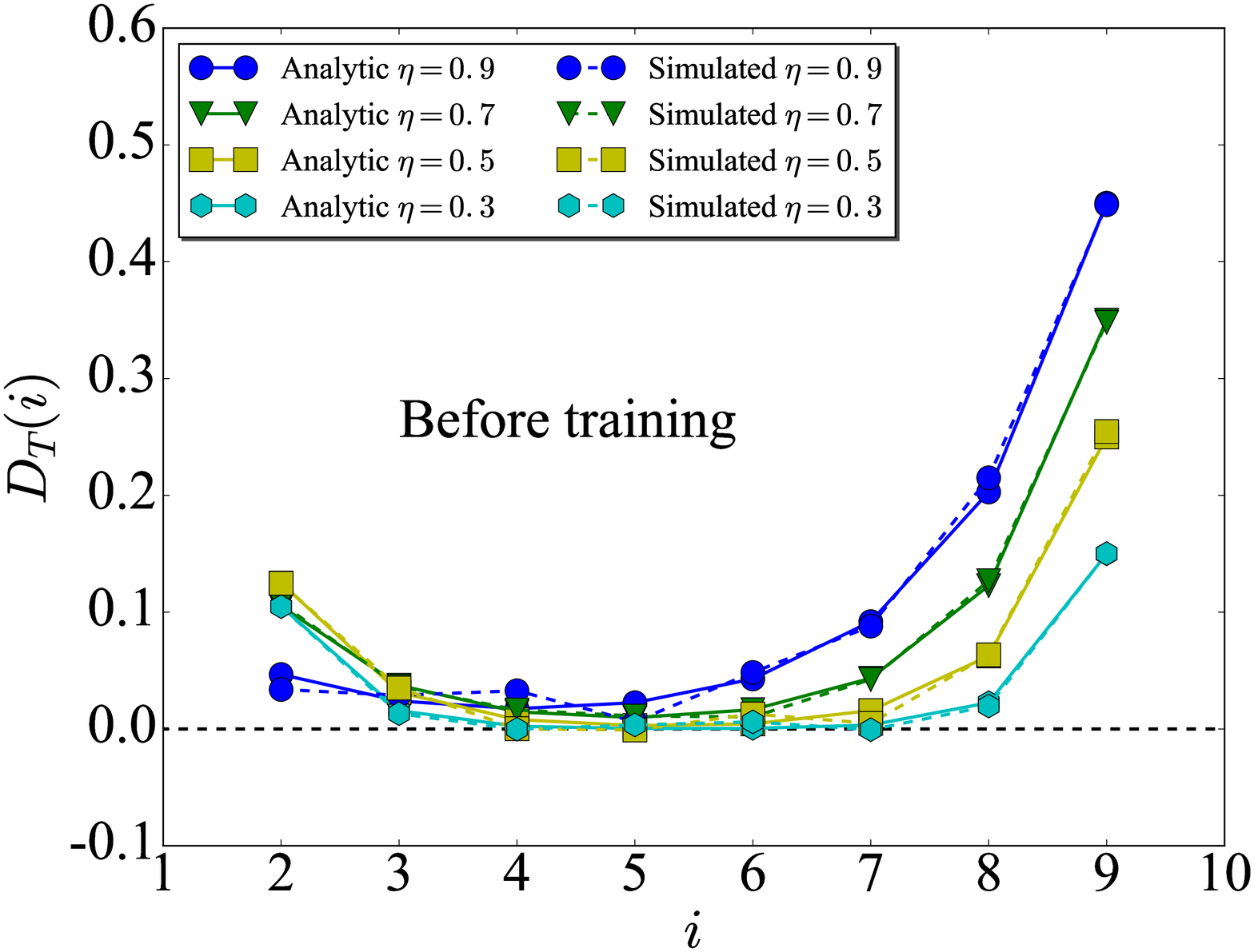}}
\subfigure[After training]{
\includegraphics[width=.28\linewidth]{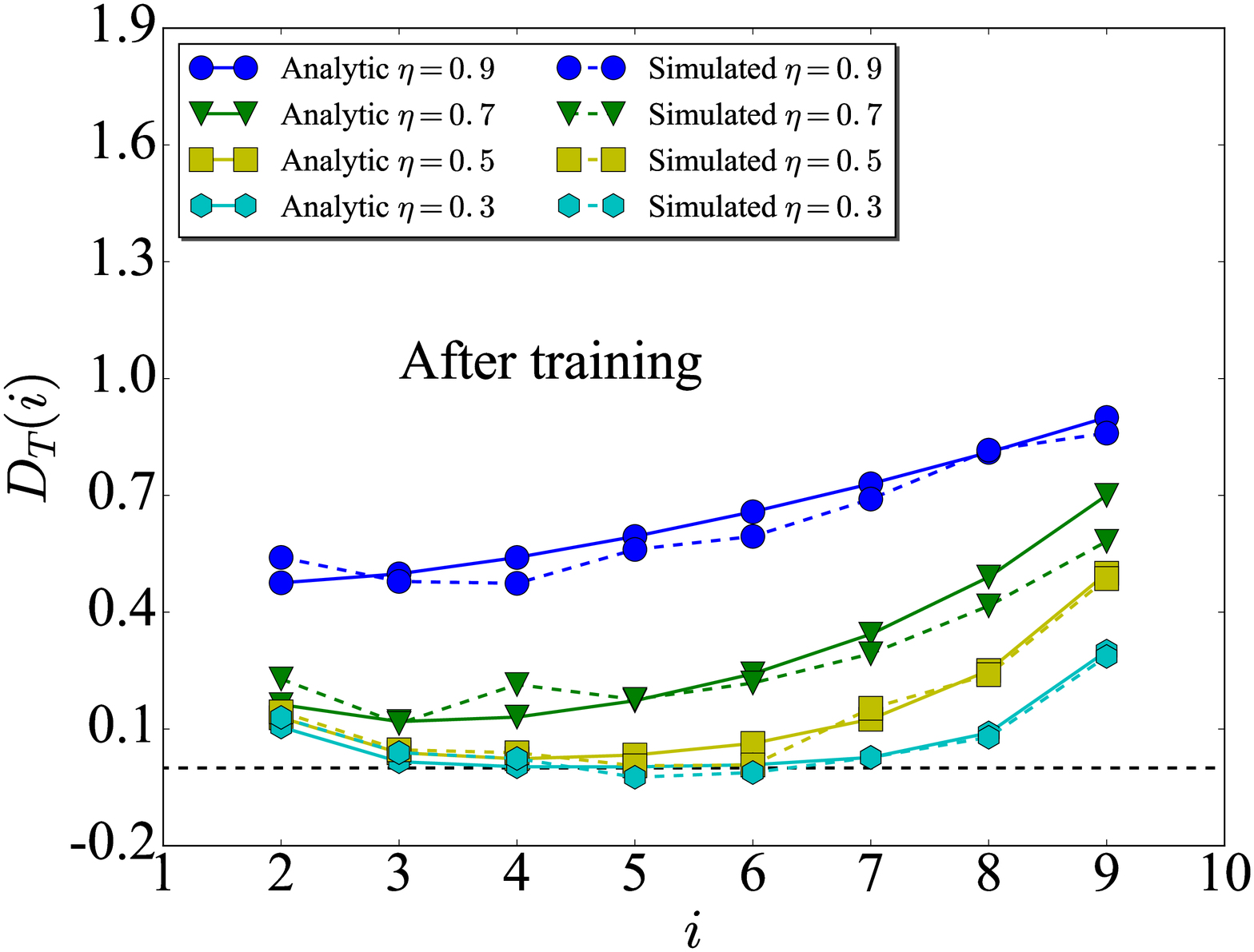}}
\subfigure[Difference]{
\includegraphics[width=.40\linewidth]{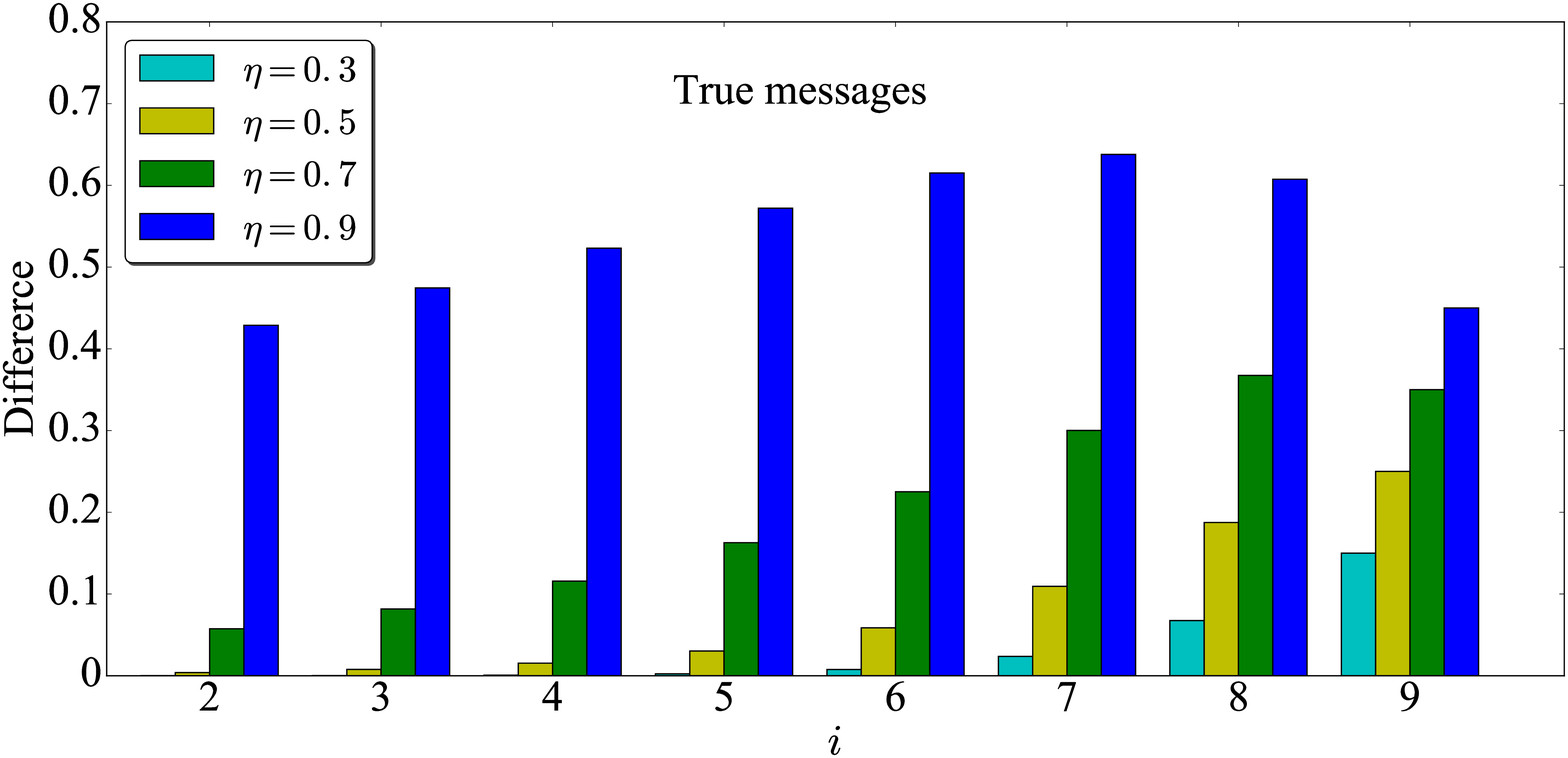}}
\caption{The analytic and simulated values of social stratification between successive nodes for the chain network as functions of the node index from one terminal as the only smart node, for the true message (a) before training, (b) after training, and (c) the difference between the two.}
\label{exp2_true}
\end{figure*}

\begin{figure*}[!t]
\centering
\subfigure[Before training]{
\includegraphics[width=.28\linewidth]{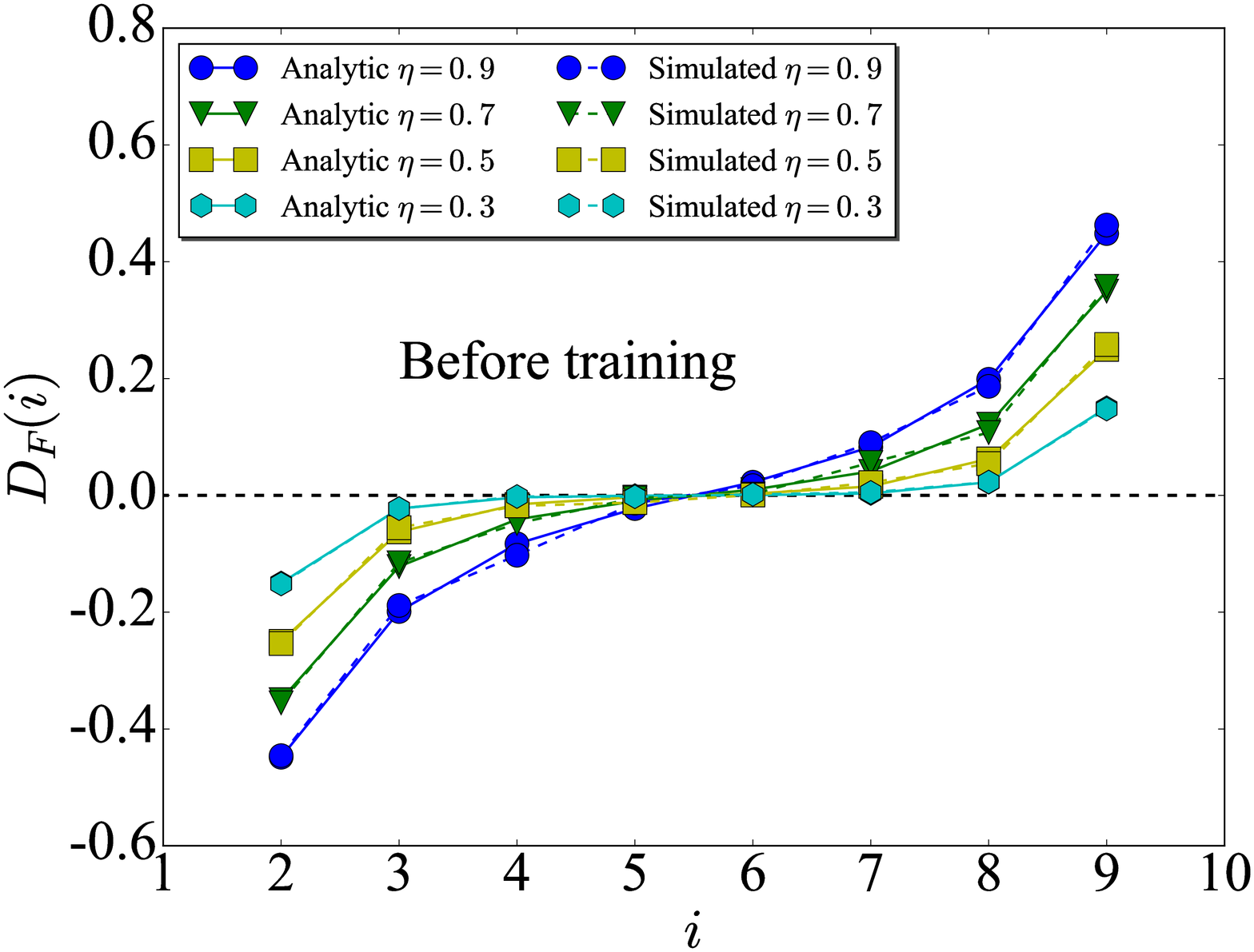}}
\subfigure[After training]{
\includegraphics[width=.28\linewidth]{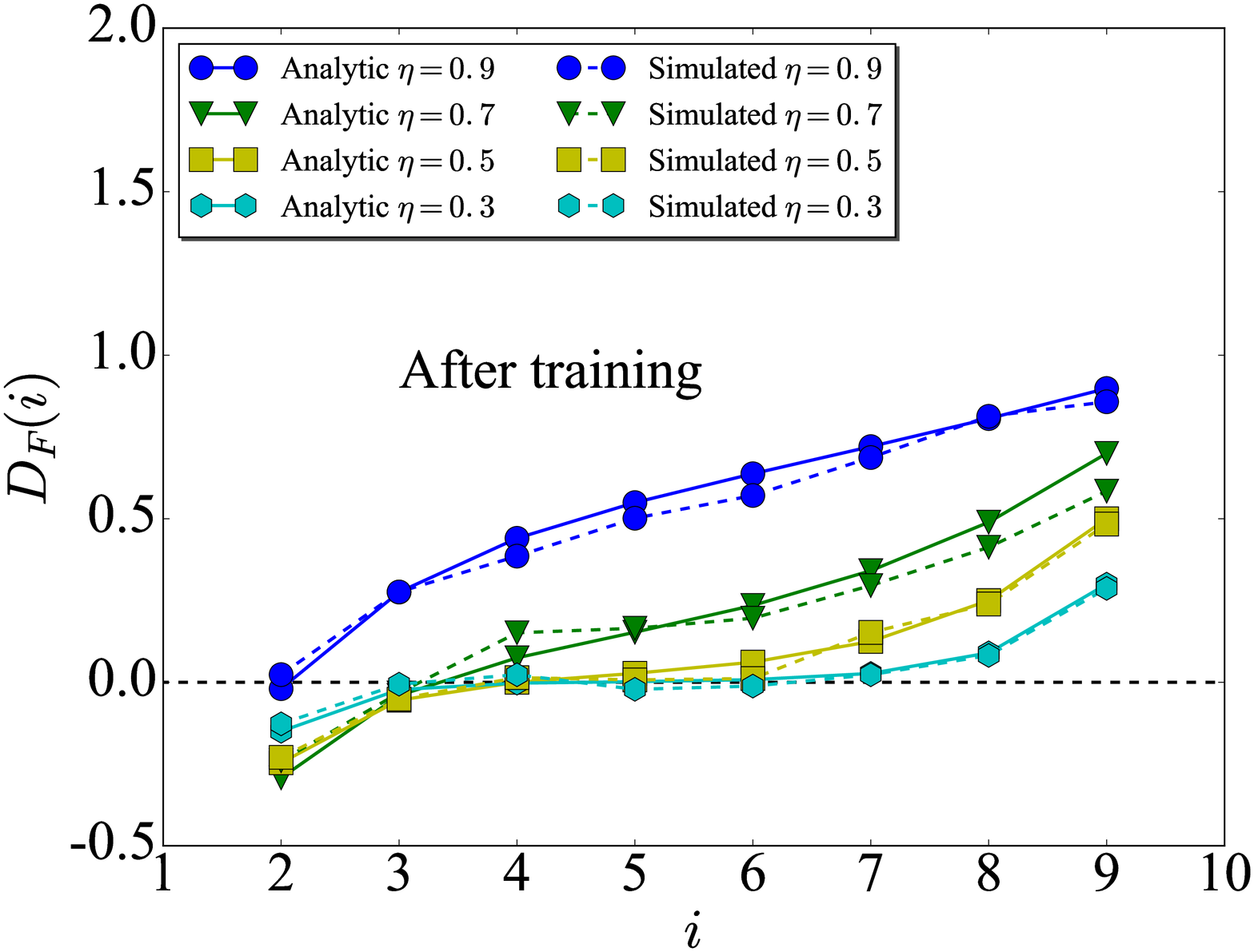}}
\subfigure[Difference]{
\includegraphics[width=.40\linewidth]{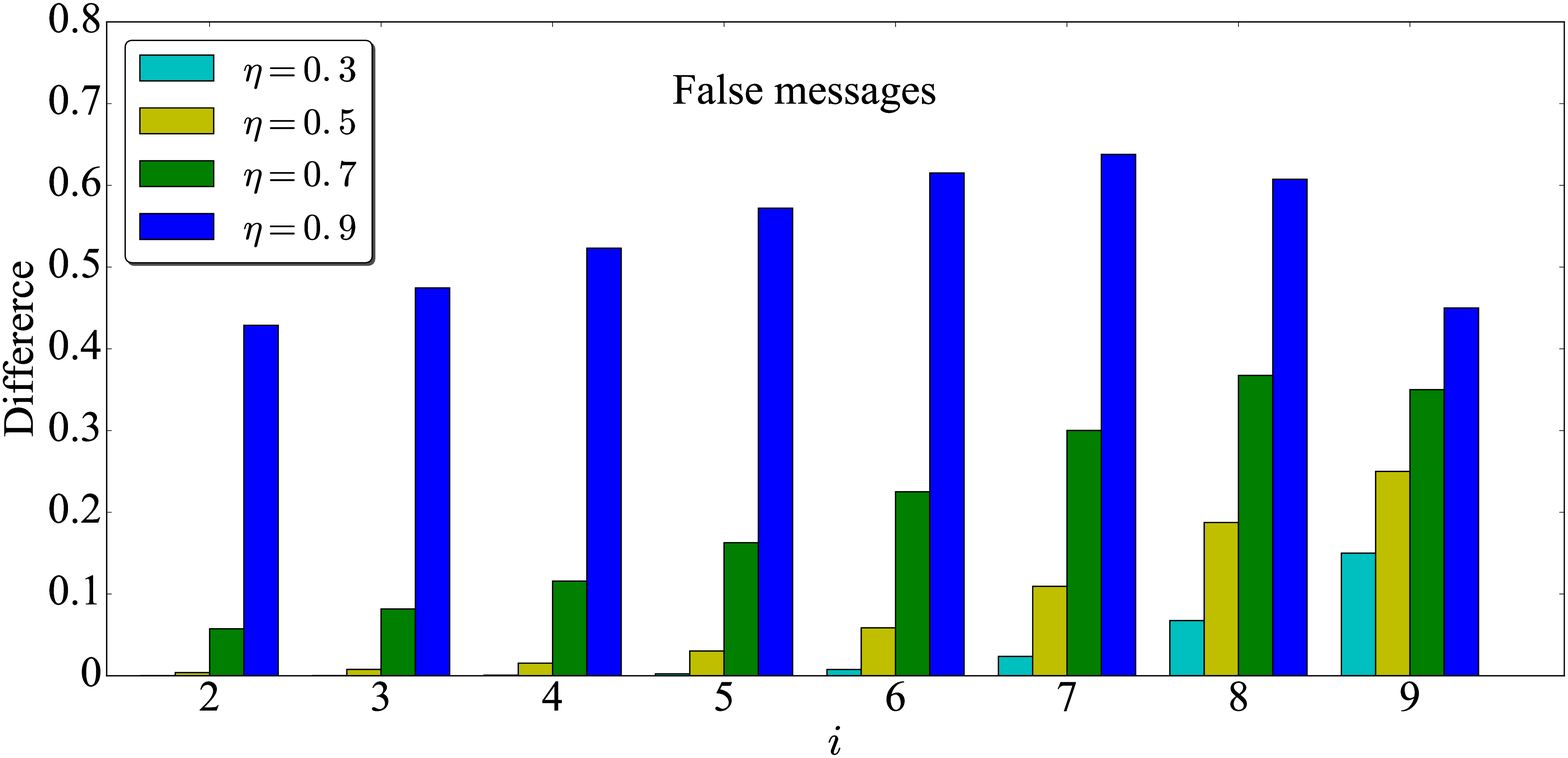}}
\caption{The analytic and simulated values of social stratification between successive nodes for the chain network as functions of the node index from one terminal as the only smart node, for the false message (a) before training, (b) after training, and (c) the difference between the two.}
\label{exp2_false}
\end{figure*}

Now, it is to compare the values of IFA before and after training, for chain and star networks, respectively. Typically, both chain and star networks have larger values of IFA after training than before, which indicates that the networks indeed become smarter, in the sense of better distinguishing true and false messages, thanks to the new self-learning mechanism introduced in Sec.~\ref{SLM}. Specifically, define the relative improvement of IFA due to the training process as
\begin{equation}
\Delta{F}=\frac{F_{A}-F_{B}}{F_{B}},
\label{Eq:RD}
\end{equation}
where $F_A$ and $F_B$ represent the values of IFA after and before training, respectively. It is found that such improvement is determined by both the network size $N$ and the NFR $\eta$, as shown in Fig.~\ref{exp1chain} (c) and Fig.~\ref{exp1star} (c). Generally, it increases as $\eta$ increase for both chain and star networks. By comparison, such improvements on star networks are much larger than those on chain networks of similar sizes. This indicates that star networks not only have a higher information filtering ability than chain networks, but also have a larger potential to be further improved by the self-learning mechanism.

\emph{\textbf{Remark 6:}} Both star and chain networks are building blocks or motifs of many real-world social networks. Our finding suggests that, by comparing with chain motifs, star motifs may play more important roles in information filtering on social networks, and such advantages may be further amplified when the network size increases, especially when the network has a self-learning ability. Thus, it seems better to let those nodes of larger degrees be smart nodes, in order to make the network have a relatively large value of IFA, since these nodes can be considered as the centers of the star motifs in the network.

\subsection{Quantifying Social Stratification}
Now, consider the social stratification in chain network, introduced by the terminal smart node. Here, a chain network of size 10 is simulated, and the information diffusion power between two successive normal nodes is compared, applying Eqs.~(\ref{Eq:DTrue}) and (\ref{Eq:DFalse}) for true and false messages, respectively, before training, and by Eqs.~(\ref{Eq:DTrueTrain}) and (\ref{Eq:DFalseTrain}) for true and false messages, respectively, after training. Similarly, in simulations, 10,000 messages are sent out from each node $v_i$ and then $D_T(i)$ and $D_F(i)$ are calculated. For the training process, set $\delta=\Delta=0.001$ and iterate 4,000,000 times.

Again, it is found that the analytic and simulated values match well in most cases, as shown in Figs.~\ref{exp2_true} (a) and (b) for the case of true message and Figs.~\ref{exp2_false} (a) and (b) for the case of false message. For the case of true message, a distinct social stratification from the smart node to the other terminal can be seen, in the sense that the normal nodes closer to the smart node have higher powers to deliver true messages, i.e., one always has $D_T(i)>0,i=2,\ldots,N-1$, for various values of NFR $\eta$, as predicted by Eqs.~(\ref{Eq:DTrue}) and (\ref{Eq:DTrueTrain}). For the case of false message, before training the switching point $i\approx{(N+1)/2}=5.5$, above which $D_F(i)>0$, while under which $D_F(i)<0$. And this switching point moves towards the smart node, i.e.,  after training it gets smaller, validating the theoretical results predicted by Eqs.~(\ref{Eq:DFalse}) and (\ref{Eq:DFalseTrain}). Moreover, it appears that the social stratification is strengthened as the value of NFR $\eta$ increases for the diffusion of both true and false messages on chain networks, before or after training. This shows that one may observe strong social stratification in a society where information can easily spread.

Next, it is to calculate the difference of social stratification between the chain network before and after training, to see whether the self-learning mechanism can enhance the social stratification. The results are shown in Fig.~\ref{exp2_true} (c) and Fig.~\ref{exp2_false} (c), for true and false messages, respectively. It is found that, for the case of true message, the social stratification between each pair of successive normal nodes is strengthened by the training process, while for the false message, the social stratification between successive normal nodes is strengthened when $i>(N+1)/2$, but is weakened or even reversed when $i<(N+1)/2$, as indicated by Eq~(\ref{Eq:HPower}). At the same time, one always has $D_F(i)>0$ as NFR $\eta\rightarrow{1}$.

\begin{figure*}[!t]
\centering
\subfigure[Before training]{
\includegraphics[width=.28\linewidth]{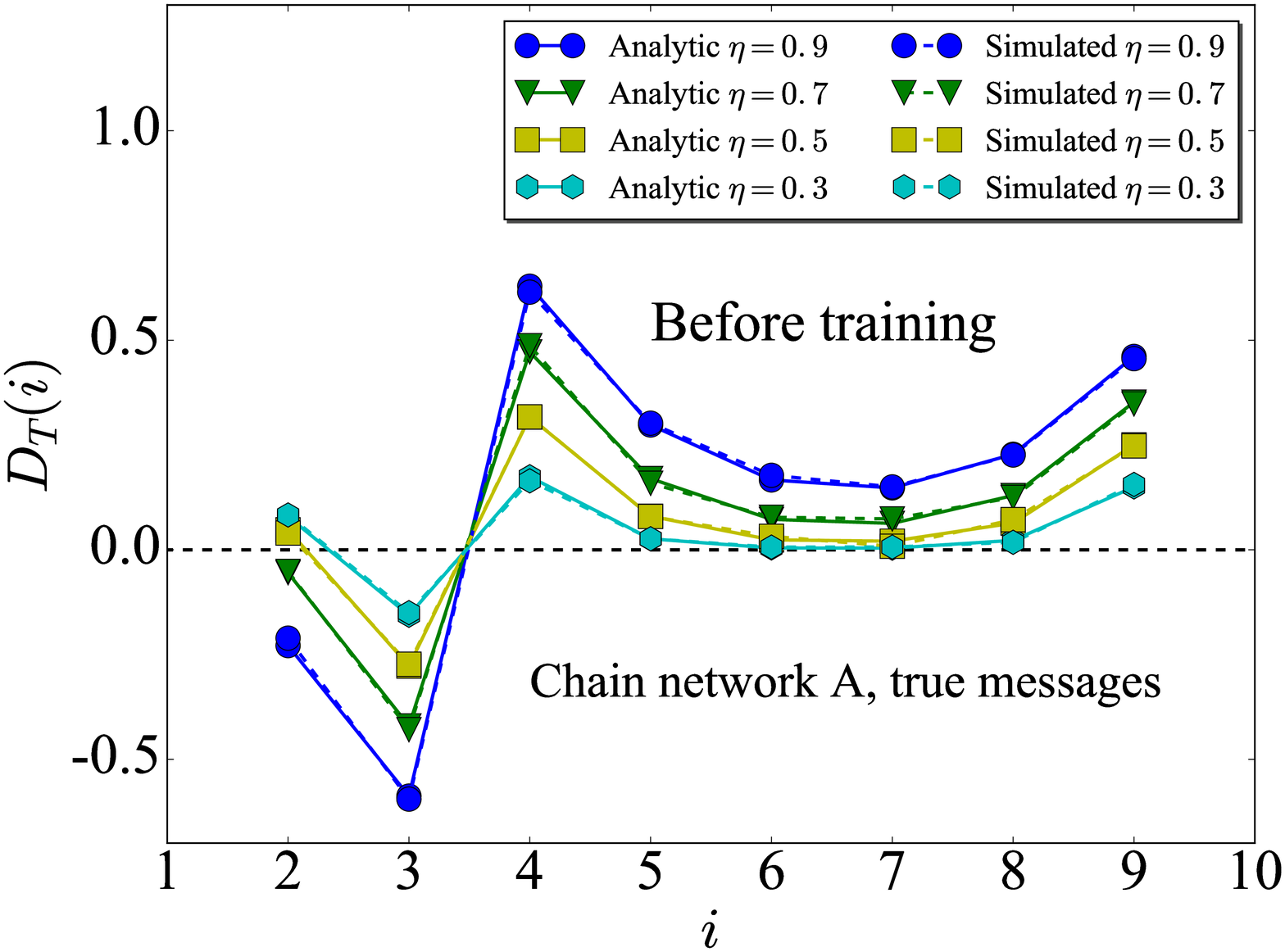}}
\subfigure[After training]{
\includegraphics[width=.28\linewidth]{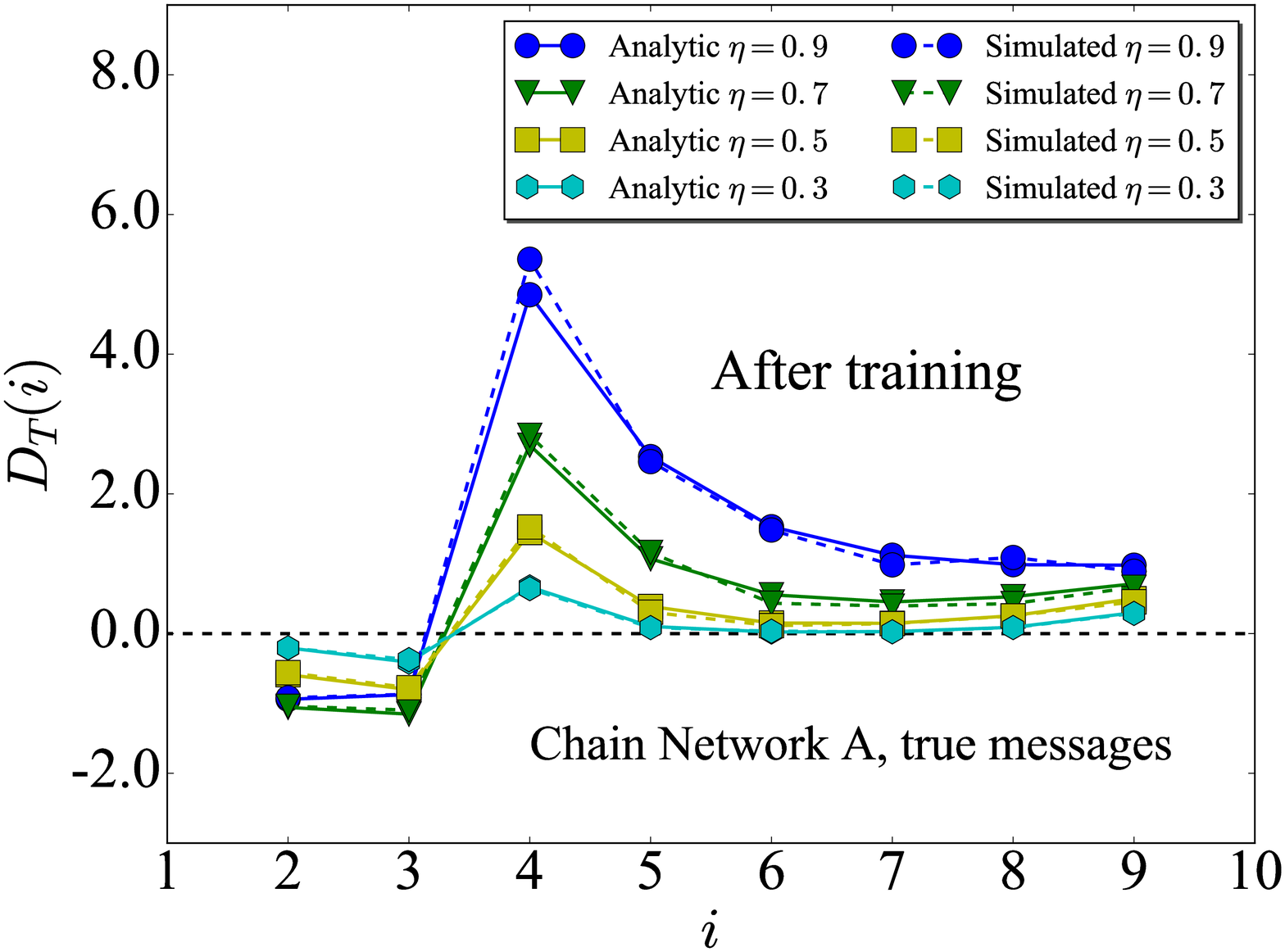}}
\subfigure[Difference]{
\includegraphics[width=.40\linewidth]{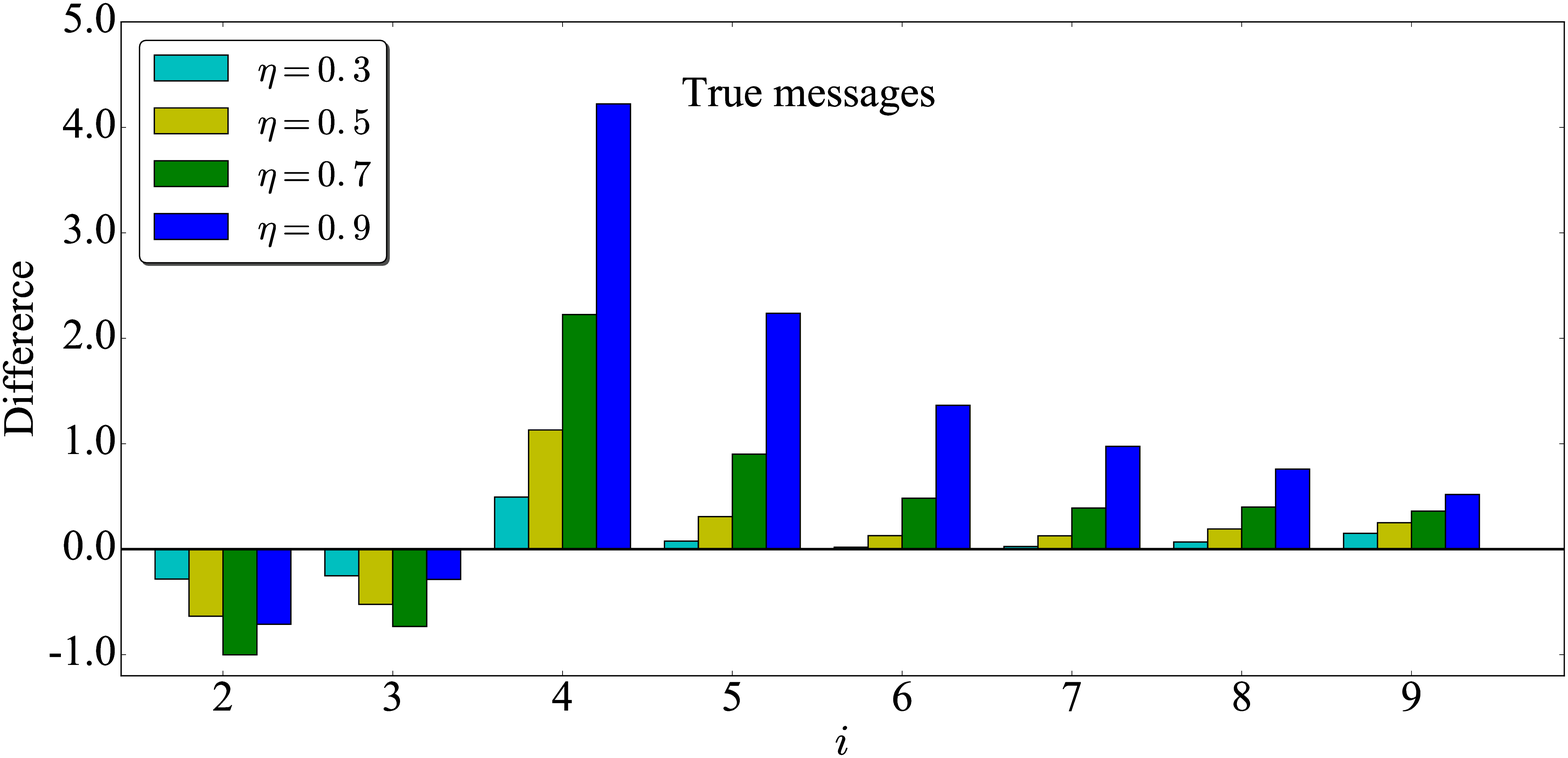}}
\caption{The analytic and simulated values of social stratification between successive nodes for chain network $A$, after it is interconnected to chain network $B$ of the same size, as functions of the node index from the smart node, for the case of true message (a) before training, (b) after training, and (c) the difference between the two. Node $v_4$ in network $A$ is interconnected to node $u_8$ in network $B$.}
\label{exp3_true}
\end{figure*}

\begin{figure*}[!t]
\centering
\subfigure[Before training]{
\includegraphics[width=.28\linewidth]{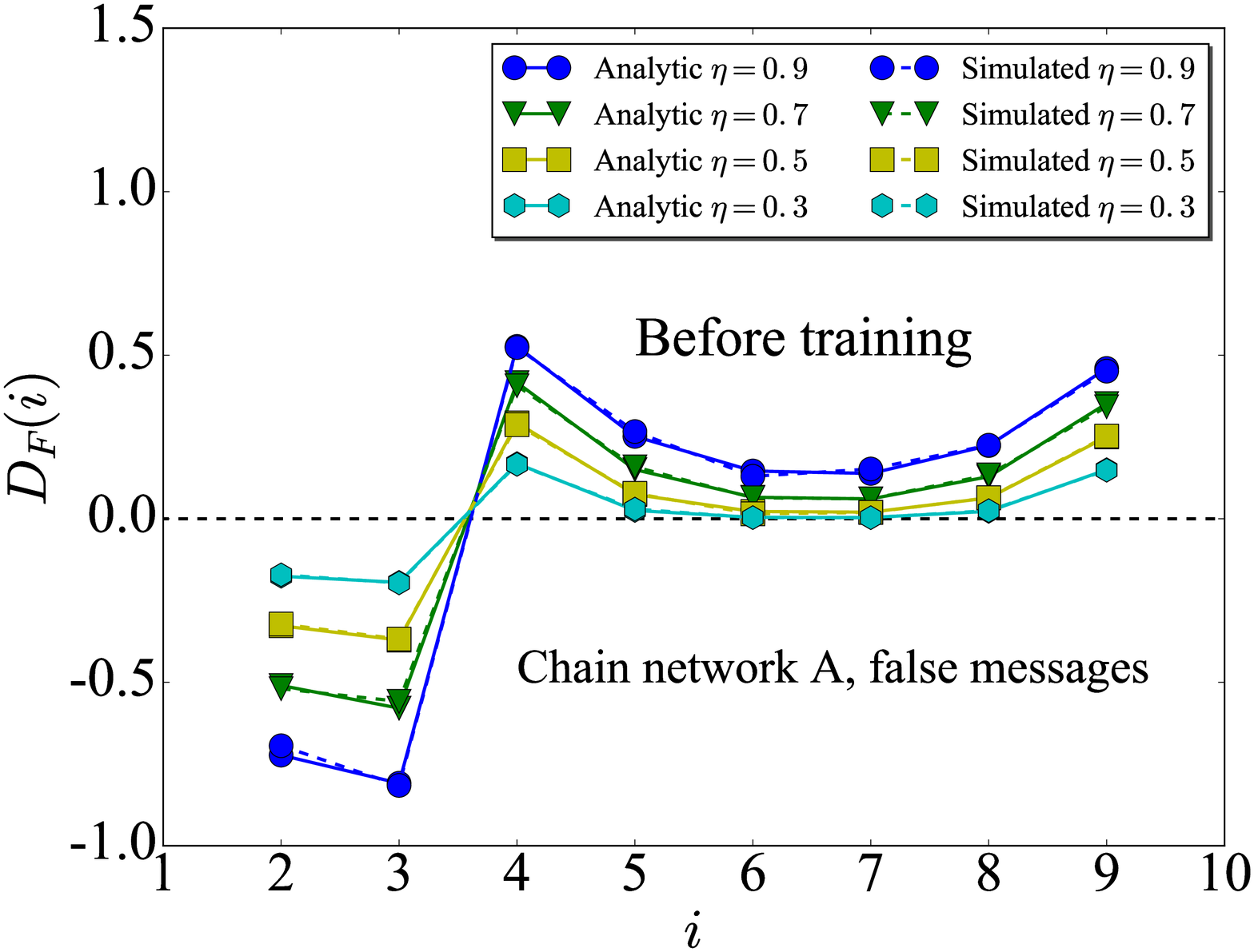}}
\subfigure[After training]{
\includegraphics[width=.28\linewidth]{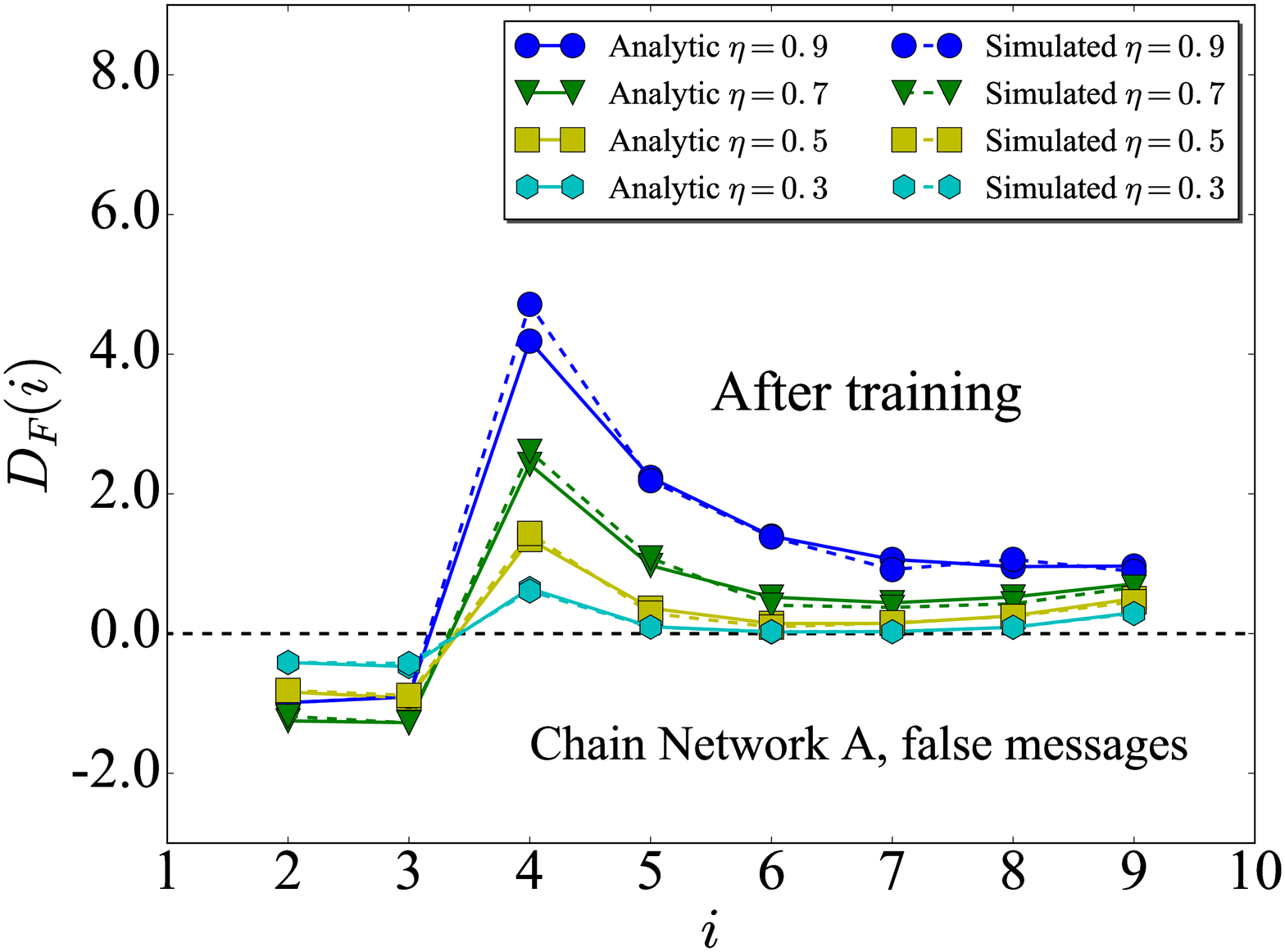}}
\subfigure[Difference]{
\includegraphics[width=.40\linewidth]{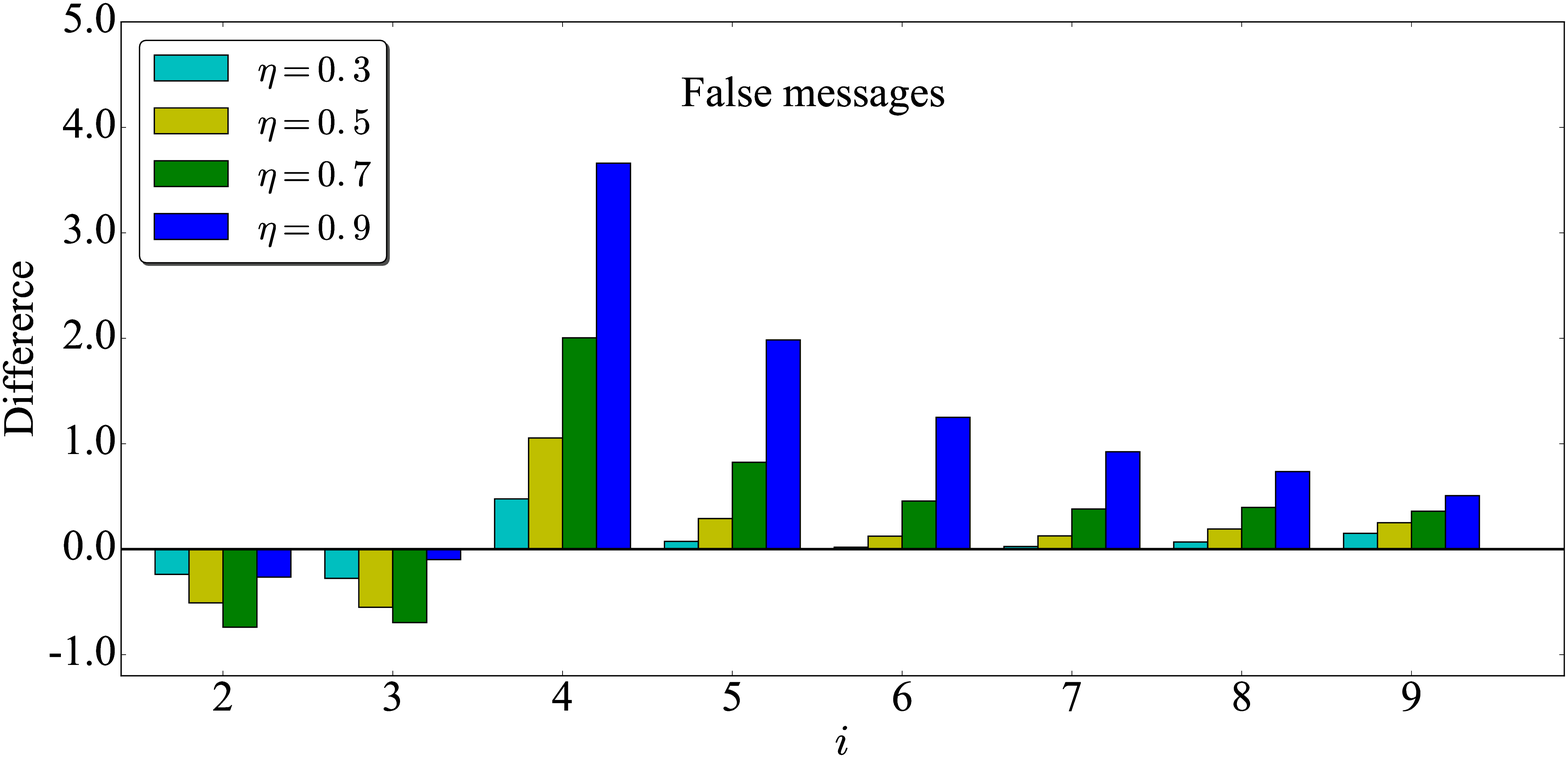}}
\caption{The analytic and simulated values of social stratification between successive nodes for chain network $A$, after it is interconnected to chain network $B$ of the same size, as functions of the node index from the smart node, for the case of false message (a) before training, (b) after training, and (c) the difference between the two. Node $v_4$ in network $A$ is interconnected to node $u_8$ in network $B$.}
\label{exp3_false}
\end{figure*}

\emph{\textbf{Remark 7:}} The above discussions are given on a chain network, since all the normal nodes in a star network are equal when the center is chosen as the only smart node. Our findings reveal a distinct social stratification from the smart node to the other terminal for the spreading of true messages on the chain network, before or after the training process. However, the normal nodes closer to the terminal nodes might have less powers to deliver false message before training, since the smart node cannot influence the diffusion of false messages directly. Interestingly, the self-learning mechanism, represented by a re-weighting process, can indeed influence the network structure, and further indirectly influence the diffusion of both true and false message. Therefore, one can find that the social stratification between each pair of successive normal nodes is strengthened by the self-learning mechanism for the case of true message, while the switching point, which determines whether the power of delivering information decreases or increases between pairwise-successive nodes, moves towards the smart node after the training process for the case of false message.

\subsection{Quantifying Crossover Effect}
In Sec.~\ref{CA}, it was theoretically proved that an interconnection between two chain networks may significantly influence the social stratification in each network. Here, simulations are performed to quantify such crossover effect and validate the theoretical results. Consider two chain networks, denoted by $A$ and $B$, each containing 10 nodes. Without loss of generality, node $v_4$ in network $A$ is interconnected to node $u_8$ in network $B$, and the social stratification in network $A$ is examined. In the simulations, 10,000 messages are sent out from each node $v_i$ and then $D_T(i)$ and $D_F(i)$ are calculated. For the training process, set $\delta=\Delta=0.001$ and iterate 8,000,000 times.

The results are shown in Figs.~\ref{exp3_true} and \ref{exp3_false}, where one can see that the analytic and simulated values match very well. By comparing Figs.~\ref{exp3_true} (a)-(b), Figs.~\ref{exp3_false} (a)-(b) with Figs.~\ref{exp2_true} (a)-(b), Figs.~\ref{exp2_false} (a)-(b), respectively, one can find that the social influence of node $v_4$ in chain network $A$, in terms of the power to deliver true or false messages, largely increases, before or after the training process, indicating a significant crossover advantage.

In order to investigate the effect of the self-learning mechanism on the crossover advantage, calculate the difference of social stratification between the chain networks before and after training, as shown in Fig.~\ref{exp3_true} (c) and Fig.~\ref{exp3_false} (c) for the cases of true and false messages, respectively. It was found that, from $v_4$ to $v_N$, the social stratification between the successive nodes is largely strengthened for both cases of true and false messages after the training process, as predicted by Eqs.~(\ref{Ineq2}) and (\ref{Ineq3}). When considering the nodes from $v_2$ to $v_4$, the situation is relatively complicated. Since for most cases, before training, i.e., $\eta=0.7,0.9$ for the true message and $\eta=0.3,0.5,0.7,0.9$ for the false message, the social stratification from $v_4$ to $v_2$ has already reversed, which is further strengthened based on the self-learning mechanism. While for the cases with $\eta=0.3,0.5$ and the true message, the social stratification between $v_2$ and $v_3$ is reversed because of the training process.

\emph{\textbf{Remark 8:}} Here, it is found that an interconnection between two chain networks can make the bridge nodes have higher social influences, in the sense of delivering more true or false messages to others, namely with crossover advantage here, while the self-learning mechanism tends to strengthen such advantage. Although in both theoretical analysis and simulations, only one of the two chain networks is discussed, one can get the same results on the other. This finding is also consistent with the theory of structural holes~\cite{burt2009structural,lou2013mining}, which suggests that individuals would benefit from filling the holes between groups that are otherwise disconnected.

\section{Conclusion\label{Conclusion}}
In this paper, we assume that the individuals in a social network have an ability to learn from historical information, based on which we propose a new information diffusion model on social networks, by considering two types of nodes, i.e., \emph{smart} and \emph{normal} nodes, and two kinds of messages, \emph{true} and \emph{false} messages, as well as a self-learning mechanism.

Based on the definition of \emph{information filtering ability} (IFA), we find that our suggested self-learning mechanism can make the network smarter, in the sense of better distinguishing true messages from the false. The introduction of a smart node causes the \emph{social stratification} in chain networks, i.e., the true messages initially posted by a node closer to the smart node can be forwarded to more other nodes. Moreover, we find that an interconnection between two chain networks can make the bridge nodes have higher social influences, in the sense of delivering more messages to others,  which is referred to as \emph{crossover advantage}. We moreover find that both social stratification and crossover advantage may be further strengthened by our proposed self-learning mechanism.

In this investigation, we focus on chain and star networks, because they are two of the most basic motifs of many real-world social networks and their simplicity also makes it feasible to theoretically analyze the information diffusion model with the self-learning mechanism. In the future, we will extend our research to some social networks of higher complexity, and further quantify the quality of messages by continuous variables, targeting more comprehensive results.


%





\ifCLASSOPTIONcaptionsoff
  \newpage
\fi



%
\bibliographystyle{IEEEtran}
\bibliography{IEEEabrv,MyRef}



%








\end{document}